\documentclass[sn-mathphys]{sn-jnl}% Math and Physical Sciences Reference Style
\jyear{2022}%

\theoremstyle{thmstyleone}%
\newtheorem{theorem}{Theorem}%  meant for continuous numbers
\newtheorem{proposition}[theorem]{Proposition}% 

\theoremstyle{thmstyletwo}%
\newtheorem{example}{Example}%
\newtheorem{remark}{Remark}%

\theoremstyle{thmstylethree}%
\newtheorem{definition}{Definition}%

\usepackage{amsfonts,amssymb,amsmath}
\usepackage{graphicx,graphics,epsfig,color}
\usepackage{multicol}
\usepackage{float}
\usepackage{subcaption}
\usepackage{comment}
\usepackage{caption}
\usepackage{hslTR}

\usepackage{url}

\newcommand{\Inc}{\textrm{Inc}}

\newcommand{\dom}{\textrm{dom}}

\newcommand{\card}{\textrm{card}}
\newcommand{\Int}{\textrm{Int}}
\newcommand{\cl}{\textrm{cl}}
\newcommand{\Single}{\textrm{Single}}

\newtheorem{lemma}{Lemma}

\begin{document}

\title[Controlled Invariants for Monotone Dynamical Systems]{Characterization, Verification and Computation of Robust Controlled Invariants for Monotone Dynamical Systems}


\author*[1,2]{\fnm{Adnane} \sur{Saoud}}\email{adnane.saoud@centralesupelec.fr}

\author[3]{\fnm{Murat} \sur{Arcak}}\email{arcak@berkeley.edu}

\affil[1]{\orgname{Laboratoire des Signaux et Syst\`emes, CentraleSup\'elec, Universit\'e Paris Saclay, Gif-sur-Yvette}, \country{France}}

\affil[2]{\orgname{College of Computing, Mohammed VI Polytechnical University, Benguerir}, \country{Morocco}}

\affil[3]{\orgdiv{Dept. of Electrical Engineering and Computer Sciences}, \orgname{University of California, Berkeley}, \country{USA}}


\abstract{In this paper, we consider the problem of computing robust controlled invariants for discrete-time monotone dynamical systems. We consider different classes of monotone systems depending on whether the sets of states, control inputs and disturbances respect a given partial order. Then, we present set-based and trajectory-based characterizations of robust controlled invariants for the considered class of systems. Based on these characterizations, we propose algorithmic approaches for the verification and computation of robust controlled invariants. Finally, illustrative examples are provided showing the merits of the proposed approach.}


\maketitle

%===============================================================================
\section{INTRODUCTION}
\label{sec:1}
 
The concept of controlled invariance plays an important role in control theory~\cite{blanchini2008set,aubin2009viability}, as it reflects the ability to control the system so that all trajectories initialized in a set remain there for all future time. This concept is crucial in safety-critical applications such as vehicle platoons~\cite{nilsson2014preliminary,smith2021monotonicity,saoud2018contract}, air traffic management~\cite{tomlin1998conflict}, robotics~\cite{wang2017safety,zanchettin2015safety} and power networks~\cite{zonetti2019symbolic} where formal proofs are required to show the ability to maintain the state in the safe region.

Different approaches have been proposed in the literature to compute controlled invariants for different classes of discrete-time systems. In~\cite{blanchini2008set}, controlled invariants are obtained as level sets of Lyapunov-like functions. Iterative algorithms are used to compute controlled invariants in~\cite{rakovic2004computation} for piecewise affine systems and more recently in~\cite{anevlavis2021controlled} for linear systems. Controlled invariants for polynomial systems have been explored using linear programming in~\cite{korda2014convex} and semidefinite programming in~\cite{sassi2012computation}. For general nonlinear systems, interval controlled invariants have been investigated recently in~\cite{saoud2021computation}. Other approaches have been proposed recently using symbolic control techniques~\cite{tabuada2009verification,saoud2019compositional}. 

In this paper, we study robust controlled invariants for discrete-time monotone dynamical systems. We consider different classes of monotone systems depending on whether the sets of states, control inputs and disturbance inputs respect a given partial order. Moreover, we focus on lower closed constraints. For the considered classes of systems and constraints, we present characterizations of the structure of the robust controlled invariants. Then we present an algorithmic procedure allowing to compute robust controlled invariants using an appropriately defined concept of feasibility. Finally, we illustrate the theoretical results on an adaptive cruise control problem.

\textbf{Related work:} The computation of controlled invariants for monotone systems has been explored for continuous time systems for the particular class of sets given by intervals. The approach in~\cite{abate2009box}  deals with monotone autonomous multi-affine systems and in~\cite{meyer2016robust} the authors present an approach to the computation of robust controlled invariants for monotone systems with inputs. In~\cite{ivanova2022lazy}, the authors use formal methods and symbolic control techniques to compute robust controlled invariants for discrete time monotone dynamical systems. The closest work to the current paper in the literature is~\cite{sadraddini2016safety} where the authors introduce a notion of s-sequence to characterize a controlled invariant for disturbance state monotone systems. Our approach generalizes the one in~\cite{sadraddini2016safety} by allowing to deal with different classes of monotone systems, by providing new characterizations of the structure of the robust controlled invariants and by using a new algorithmic procedure to compute robust controlled invariants based on tools from multidimensional binary search algorithms used in multi-objective optimization~\cite{legriel2010approximating}.

A preliminary version of this work is currently under review in the IEEE Conference on Decision and Control (CDC) 2022~\cite{saoud2022}. In the current paper we are providing proofs for different results, which has not been done in the conference version. Moreover, while in~\cite{saoud2022}, we are only dealing with the computation of the controlled invariants, we are also presenting here another algorithm for the verification of controlled invariants together with a new numerical example.

The remainder of this paper is organized as follows. In Section~\ref{sec:3} we introduce the class of systems we consider. Section~\ref{sec:4} introduces the concept of robust controlled invariants. In Section~\ref{sec:5} and~\ref{sec:6}, we present different characterizations of robust controlled invariants. Section~\ref{sec:7} presents algorithms to verify and compute controlled invariants. Finally, Section~\ref{sec:5} presents numerical results validating the merits of the proposed approach. To improve the readability of the paper, all the proofs are given in the appendix.

\paragraph{Notation} 
The symbols $\mathbb{N}$, $ \mathbb{N}_{>0} $, $\mathbb{R}$ and $\mathbb{R}_{>0}$ denote the set of positive integers, non-negative integers, real and non-negative real numbers, respectively. Given $N \in \mathbb{N}_{>0}$ and a set $Y \subseteq \mathbb{R}^n$, $Y^w$ denotes the set of infinite sequences of elements of $Y$. For a map $f:\mathbb{R}^n \rightarrow \mathbb{R}^m $, $\dom{f}:=\{x \in \mathbb{R}^n: f(x) \text{ is well defined}\}$. Given a nonempty set $K$, $\Int(K)$ denotes it interior, $\cl(K)$ denotes its closure, $\partial K$ denotes its boundary and $\overline{K}$ its complement. For a set $K$, the operator $\Single(K)$ randomly selects a unique element from the set $K$. The Euclidean norm is denoted by $\|.\|$. For $x \in \mathbb{R}^n$ and for $\varepsilon >0$, $\mathcal{B}_{\varepsilon}(x)=\{z \in \mathbb{R}^n \mid \|z-x\| \leq \varepsilon \}$ and for a set $K \subseteq \mathbb{R}^n$, $\mathcal{B}_{\varepsilon}(K)= \cup_{x \in K}\mathcal{B}_{\varepsilon}(x)$. 
\section{Preliminaries}
\label{sec:2}
\subsection{Partial orders}
A partially ordered set $\mathcal{L}$ has an associated binary relation $\leq_{\mathcal{L}}$ where for all $l_1,l_2,l_3 \in \mathcal{L}$, the binary relation satisfies: (i) $l_1 \leq_{\mathcal{L}} l_1 $, (ii) if $l_1 \leq_{\mathcal{L}} l_2$ and $l_2 \leq_{\mathcal{L}} l_1$ then $l_1=_{\mathcal{L}}l_2$ and, (iii) if $l_1 \leq_{\mathcal{L}} l_2$ and $l_2 \leq_{\mathcal{L}} l_3$ then $l_1 \leq_{\mathcal{L}} l_3$. If neither $l_1 \leq_{\mathcal{L}} l_2$ nor $l_2 \leq_{\mathcal{L}} l_1$ holds, we say that $l_1$ and $l_2$ are incomparable. The set of all incomparable couples in $\mathcal{L}$ is denoted by $\Inc_{\mathcal{L}}$. We say that $l_1<_{\mathcal{L}} l_2$ iff $l_1\leq_{\mathcal{L}} l_2$ and $l_1\neq_{\mathcal{L}} l_2$. Similarly, a partial ordering $m \leq_{\mathcal{L}^w} n$ between a pair of infinite sequences $m=m_0m_1\ldots$ and $n=n_1n_2\ldots$ holds if and only if $m_k \leq_{\mathcal{L}} n_k $ for all $k \in \mathbb{N}_{\geq 0}$. 

For a partially ordered set $\mathcal{L}$, closed intervals are $[x,y]_{\mathcal{L}}:=\{z \mid x \leq_{\mathcal{L}} z \leq_{\mathcal{L}} y\}$. Given a partially ordered set $\mathcal{L}$, for $a\in \mathcal{L}$ let $\downarrow a :=\{x \in \mathcal{L} \mid x \leq_{\mathcal{L}} a \}$ and $\uparrow a :=\{x \in \mathcal{L} \mid a \leq_{\mathcal{L}} x \}$. When $A \subseteq \mathcal{L}$ then its lower closure (respectively upper closure) is $\downarrow A :=\bigcup_{a \in A} \downarrow a $ (respectively $\uparrow A :=\bigcup_{a \in A} \uparrow a $). A subset $A \subseteq \mathcal{L}$ is said to be {\it lower-closed} (respectively {\it upper-closed}) if $\downarrow A = A$ (respectively $\uparrow A = A$). We have the following definitions relative to partially ordered sets.
\begin{definition}
Let $\mathcal{L}$ be a partially ordered set and $A\subseteq \mathcal{L}$. The set $A$ is said to be {\it bounded below} (in $\mathcal{L}$) if there exists a compact set $B \subseteq \mathcal{L}$ such that $A \subseteq \uparrow B$. Similarly, the set $A$ is said to be {\it bounded above} (in $\mathcal{L}$) if there exists a compact set $B \subseteq \mathcal{L}$ such that $A \subseteq \downarrow B$.
\end{definition}
\begin{definition}
\label{def:min}
    Let $\mathcal{L}$ be a partially ordered set and consider a closed subset $A\subseteq \mathcal{L}$. If the set $A$ is bounded below then the set of {\it minimal elements} of $A$ is defined as
    $\min(A):=\{x \in A \mid \forall x_1 \in A,\, x \leq_\mathcal{L} x_1 \text{ or } (x,x_1)\in \Inc_\mathcal{L}\}$. Similarly, if the set $A$ is bounded above then the set of {\it maximal elements} of $A$ is defined as $\max(A) := \{x \in A \mid \forall x_1 \in A,\, x \geq_\mathcal{L} x_1 \text{ or } (x,x_1)\in \Inc_\mathcal{L}\}$.
\end{definition}

\begin{figure}[!t]
	\begin{center}
		\includegraphics[scale=0.5]{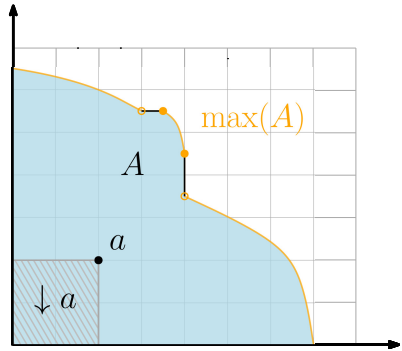}\\
	\end{center}
	\caption{A lower closed set $A \subseteq \mathbb{R}_{\geq 0}^2$, with the standard ordering, in blue. The set of its maximal elements $\max(A)$ is presented in orange. The lower closure of a point $a \in A$ is presented in dashed gray.}
	\label{fig:lower_closed}	
\end{figure}

An illustration of the concepts of lower-closed sets and maximal elements is provided in Figure~\ref{fig:lower_closed}. It was shown in~\cite{davey2002introduction} that lower and upper-closed sets satisfy the following property.

\begin{proposition}
\label{pro:closed}
    Let $\mathcal{L}$ be a partially ordered set and consider a collection of subsets $A_i\subseteq \mathcal{L}$, $i\in \{1,2,\ldots,p\}$. The following holds:
    \begin{itemize}
        \item[(i)] If for all $i \in \{1,2,\ldots,p\}$, $A_i$ is lower closed then $\cup_{i=1}^p A_i$ and $\cap_{i=1}^p A_i$ are lower closed;
        
        \item[(i)] If for all $i \in \{1,2,\ldots,p\}$, $A_i$ is upper closed then $\cup_{i=1}^p A_i$ and $\cap_{i=1}^p A_i$ are upper closed. 
    \end{itemize}
\end{proposition}
In the rest of the paper, we will focus on lower-closed sets; analogous results can be formulated for upper-closed sets.

% \textcolor{blue}{For a lower-closed sets, we use the operator $\max$ to introduce the notion of the  basis~\cite{finkel2001well}, which serves as a simpler representation of the lower-closed set. 
% \begin{definition}
% 	\label{def:basis}
% 	Let $\mathcal{L}$ be a finite partially ordered set. Let $Z \subseteq\mathcal{L}$ be a lower-closed set. A set $B=\{s_1,\ldots,s_N\} \subseteq Z$ is said to be the basis of $Z$, denoted $B=\Bas(Z)$, if $B=\max(Z)$ or in other words
% 	\begin{itemize}
% 	    \item $Z=\bigcup_{i= 1,\ldots,N}\downarrow s_i$;
% 	    \item for all $s_i,s_j \in B$, if $s_i \neq s_j$ then $(s_i,s_j) \in \Inc_\mathcal{L}$.
% 	\end{itemize}
% \end{definition}}
% The existence and uniqueness of a finite basis of a finite lower-closed set follow from the fact that the relation $\leq_\mathcal{L}$ is a well-quasi-order~\cite{higman1952ordering}. An illustration of the concept of basis is given in Figure~\ref{fig:bas}.}

\subsection{Continuity of Set-Valued Maps} 

In this section, we recall the following continuity notions for set-valued maps~\cite{aubin2009set}. 

\begin{definition} \label{deflusc}
Consider a set-valued map $F: \mathcal{X} \rightrightarrows \mathbb{R}^n$, where $\mathcal{X} \subset \mathbb{R}^m$ and $F(x)$ is compact for all $x \in \mathcal{X}$.
\begin{itemize}
\item The map $F$ is said to be \textit{lower semicontinuous} at $x \in \mathcal{X}$ if for each 
$\epsilon > 0$ and $y_x \in F(x)$, there exists $\eta >0 $ such that the following property holds: for each $z \in \mathcal{B}_{\eta}(x) \cap \mathcal{X}$, there exists $y_z \in F(z)$ such that $\|y_z - y_x\| \leq \epsilon$;  
\item The map $F$ is said to be \textit{upper semicontinuous} at $x \in \mathcal{X}$ if, for each 
$\epsilon > 0$, there exists $\eta >0$ such that $F(\mathcal{B}_{\eta}(x)) \cap \mathcal{X}) \subset \mathcal{B}_{\varepsilon}(F(x))$;
\item The map $F$ is said to be \textit{continuous} at $x \in \mathcal{X}$ if it is both upper and lower semicontinuous at $x$.
\item The map $F$ is said to be lower, upper semicontinuous, or continuous if, respectively, it is lower, upper semicontinuous, or continuous for all $x \in \mathcal{X}$.
\item For $L \geq 0$, the set valued map $F$ is said to be {\it $L$-Lipschitz} if for all $x_1,x_2 \in \mathcal{X}$, $F(x_1) \subseteq \mathcal{B}_{L\| x_1-x_2 \|}(F(x_2))$.
\end{itemize}
\end{definition}

\subsection{Discrete-time control systems}

In this paper, we consider the class of discrete-time control systems $\Sigma$ of the form:
\begin{equation}
\label{dis_sys}
x(k+1) = f(x(k),u(k),d(k))
\end{equation}
where $x(k)\in \mathcal{X}$ is a state, $u(k) \in \mathcal{U}$ is a control input and $d(k)\in \mathcal{D}$ is a disturbance input. The trajectories of (\ref{dis_sys}) are denoted by $\Phi(.,x_0,\mathbf{u},\mathbf{d})$ where $\Phi(k,x_0,\mathbf{u},\mathbf{d})$ is the state reached at time $k \in \mathbb{N}_{\geq 0}$ from the initial state $x_0$ under the control input $\mathbf{u}:\mathbb{N}_{\geq 0} \rightarrow \mathcal{U}$ and the disturbance input $\mathbf{d}:\mathbb{N}_{\geq 0} \rightarrow \mathcal{D}$. For $X \subseteq \mathcal{X}$, $U \subseteq \mathcal{U}$ and $D \subseteq \mathcal{D}$, we use the notation $f(X,U,D)=\{f(x,u,d) \mid x\in X,~u\in U,~d\in D\}$. 

When the control inputs of system (\ref{dis_sys}) are generated by a state-feedback controller $\kappa:\mathcal{X} \rightarrow \mathcal{U}$,  the dynamics of the closed-loop system is given by
\begin{equation}
\label{eqn:syst_cl}
    x(k+1) = f(x(k),\kappa(x(k)),\mathbf{d}(k))
\end{equation}
and its trajectories are denoted by $\Phi_\kappa(.,x_0,\mathbf{d})$. By abuse of notation, in the rest of the paper we use $\Phi_\kappa(.,x_0,D)$ to denote $\{\Phi_\kappa(.,x_0,\mathbf{d})\mid \mathbf{d}:\mathbb{R}_{\geq 0} \rightarrow D\}$.

\section{Monotone control systems}
\label{sec:3}
In this section, we introduce classes of monotone discrete-time control systems, that preserve order with respect to states, disturbance inputs and control inputs. Then, we provide characterizations of the considered classes of systems.

\begin{definition}
Consider the discrete-time control system $\Sigma$ in (\ref{dis_sys}).
The system $\Sigma$ is said to be:
\begin{itemize}
    \item {\it State monotone (SM)} if its set of states is equipped with a partial order $\leq_{\mathcal{X}}$, and for all $x_1,x_2 \in \mathcal{X}$, for all $u \in \mathcal{U}$ and for all $d\in \mathcal{D}$, if $x_1 \leq_{\mathcal{X}} x_2$ then  $f(x_1,u,d) \leq_{\mathcal{X}} f(x_2,u,d)$;
    \item {\it Control-state monotone (CSM)} if its sets of states and control inputs are equipped with partial orders $\leq_{\mathcal{X}}$ and $\leq_{\mathcal{U}}$, respectively, and for all $x_1,x_2 \in \mathcal{X}$, for all $u_1,u_2 \in \mathcal{U}$ and for all $d \in \mathcal{D}$, if $x_1 \leq_{\mathcal{X}} x_2$ and $u_1\leq_{\mathcal{U}} u_2$ then $f(x_1,u_1,d) \leq_{\mathcal{X}} f(x_2,u_2,d)$;
    \item {\it Disturbance-state monotone (DSM)} if its sets of states and distrubance inputs are equipped with partial orders $\leq_{\mathcal{X}}$ and $\leq_{\mathcal{D}}$, respectively, and for all $x_1,x_2 \in \mathcal{X}$, for all $u \in \mathcal{U}$ and for all $d_1,d_2\in \mathcal{D}$, if $x_1 \leq_{\mathcal{X}} x_2$ and $d_1\leq_{\mathcal{D}} d_2$ then $f(x_1,u,d_1) \leq_{\mathcal{X}} f(x_2,u,d_2)$;
    \item {\it Control-disturbance-state monotone (CDSM)} if its sets of states, inputs and disturbances are equipped with partial orders, $\leq_{\mathcal{X}}$, $\leq_{\mathcal{U}}$ and $\leq_{\mathcal{D}}$, respectively, and for all $x_1,x_2 \in \mathcal{X},\ u_1,u_2 \in \mathcal{U}$ and for all $d_1,d_2\in \mathcal{D}$, if $x_1 \leq_{\mathcal{X}} x_2$, $u_1 \leq_{\mathcal{U}} u_2$ and $d_1 \leq_{\mathcal{D}} d_2$ then $ f(x_1,u_1, d_1)\leq_{\mathcal{X}} f(x_2,u_2,d_2)$.
\end{itemize}
\end{definition}
\begin{remark}
In this paper, different types of monotonicity are defined with respect to the state, control input and distrubance input. The SM, (DSM and CDSM, respectively) properties defined in this paper correspond to the discrete-time versions of the concept of monotonicity in~\cite{smith2008monotone} (\cite{angeli2003monotone} and~\cite{meyer2015adhs}, respectively). 
\end{remark}
From the definitions above, it can be seen that a CDSM system is a CSM and DSM system, and that a DSM or CSM is a SM system. The notions above can be easily verified via the Kamke-Muller conditions~\cite{smith2008monotone} for continuously differentiable vector fields as follows: The system $\Sigma$ in (\ref{dis_sys}) with $x(k) \in \mathcal{X} \subseteq \mathbb{R}^n$, $u(k) \in \mathcal{U} \subseteq \mathbb{R}^m$ and $d(k) \in \mathcal{D} \subseteq \mathbb{R}^p$ is 
\begin{itemize}
    \item SM if $\frac{\partial f_i}{\partial x_j} \geq 0$ for all $i,j \in \{1,2,\ldots,n\}$;
     \item CSM if $\frac{\partial f_i}{\partial x_j} \geq 0$ and $\frac{\partial f_i}{\partial u_h} \geq 0$ for all $i,j \in \{1,2,\ldots,n\}$ and for all $h\in \{1,2,\ldots,m\}$;
    \item DSM if $\frac{\partial f_i}{\partial x_j} \geq 0$ and $\frac{\partial f_i}{\partial d_h} \geq 0$ for all $i,j \in \{1,2,\ldots,n\}$ and for all $h\in \{1,2,\ldots,p\}$;
    \item CDSM if $\frac{\partial f_i}{\partial x_j} \geq 0$,  $\frac{\partial f_i}{\partial u_h} \geq 0$ and $\frac{\partial f_i}{\partial d_l} \geq 0$ for all $i,j \in \{1,2,\ldots,n\}$, for all $h\in \{1,2,\ldots,m\}$ and for all $l \in \{1,2,\ldots,p\}$.
\end{itemize}
where $\geq 0$ is the usual total order on $\mathbb{R}$.

The following examples illustrate the difference between the different versions of monotonicity introduced above.

\begin{example} We present examples of the considered classes of systems:
\begin{itemize}
    \item The system described by 
$$x(k+1)=x(k)+u(k)d(k)\sin(u(k)d(k))$$
with $x(k),u(k),d(k) \in \mathbb{R}$, is SM without being DSM nor CDSM for the usual total order on $\mathbb{R}$.

\item The system described by
\begin{equation*}
		x(k+1)=\;
		\left\{
		\begin{array}{c c}
 		A_1x(k)+d(k) \;\text{ if }\; u=1  \\
		A_2x(k)+d(k) \;\text{ if }\; u=2
		\end{array}
		\right. 
\end{equation*}
		with $x(k),d(k) \in \mathbb{R}^2$ and $u(k) \in \{1,2\}$,  $A_1=\begin{pmatrix} 0.8 & 0.1 \\ 2 & 4 \end{pmatrix}$ and $A_2=\begin{pmatrix} 5 & 0.2 \\ 8 & 0 \end{pmatrix}$, is DSM without being CSM nor CDSM for the usual partial order on $\mathbb{R}^2$;

\item The system described by 
$$x(k+1)=x(k)+u(k)+d(k)$$
with $x(k),u(k),d(k) \in \mathbb{R}$, is CDSM for the usual total order on $\mathbb{R}$.
\end{itemize}

\end{example}

\begin{remark}
{Monotone systems constitute a broad class of systems that arise in biology~\cite{angeli2003monotone}, traffic flow models~\cite{coogan2017formal}, microgrids~\cite{zonetti2019decentralized}, and other applications. Monotonicity may appear to be restrictive when judged from the sign conditions for monotonicity with respect to the positive orthant, but graphical tests exist to detect monotonicity with respect to other orthants and can be modified with changes of variables to detect monotonicity with respect to broader cones. In addition, physical insights, e.g. restricting the dynamics to hyperplanes associated with conservation laws, may reveal monotonicity that is not apparent; this has been observed in biomolecular signaling cascades. Finally, algorithms exist to decompose systems into monotone components \cite{dasgupta2007algorithmic}.}
\end{remark}

Next we give an auxiliary lemma, which enables us to present equivalent characterizations of the proposed classes of monotone systems.

\begin{lemma}
	\label{lem:mono}
	Let $\mathcal{L}$ be a partially ordered set and $A,B \subseteq \mathcal{L}$. We have $A\subseteq \downarrow B$ if and only if for any $a\in A$, there exists $b\in B$ such that $a \leq_{\mathcal{L}} b$.
\end{lemma}
The proof follows immediately from the fact that $\downarrow B= \{z \in\mathcal{L} \mid \exists\,b\in B \text{ satisfying } z \leq_{\mathcal{L}} b \}$.

\begin{proposition}
\label{prop:characterizations_monotone}
Consider the discrete-time control system $\Sigma$ in (\ref{dis_sys}), the following properties hold:
\begin{itemize}
    \item[(i)] The system $\Sigma$ is SM if and only if for all $x \in X$, $u \in U$ and $d \in D$ we have
    $$f(\downarrow x,u,d) \subseteq \downarrow f(x,u,d)$$
    
    \item[(ii)] The system $\Sigma$ is CSM if and only if for all $x \in X$, $u \in U$ and $d \in D$ we have
    $$f(\downarrow x,\downarrow u, d) \subseteq \downarrow f(x,u,d)$$
    
    \item[(iii)] The system $\Sigma$ is DSM if and only if for all $x \in X$, $u \in U$ and $d \in D$ we have
    $$f(\downarrow x, u,\downarrow d) \subseteq \downarrow f(x,u,d)$$
    
    \item[(iv)] The system $\Sigma$ is CDSM if and only if for all $x \in X$, $u \in U$ and $d \in D$ we have
    $$f(\downarrow x,\downarrow u,\downarrow d) \subseteq \downarrow f(x,u,d)$$
\end{itemize}
\end{proposition}

The following auxiliary result characterizes  the monotonicity property of the closed loop controlled system.

\begin{lemma}
\label{lem:cont_mono}
Consider the system $\Sigma$ in (\ref{dis_sys}). 
If the system $\Sigma$ is CDSM and if the controllers $\kappa_1,\kappa_2:\mathcal{X}\rightarrow \mathcal{U}$ satisfy 
\begin{equation}
\label{eqn:cont_mono}
    \kappa_1(x_1) \leq_{\mathcal{U}} \kappa_2(x_2),~~ \forall x_1,x_2 \in \mathcal{X}, \text{ with } x_1 \leq_{\mathcal{X}} x_2,
\end{equation}
then for all $x_1^0,x_2^0 \in X$, with $x_1^0 \leq_{\mathcal{X}} x_2^0$, and for all $\mathbf{d}_1,\mathbf{d}_2:\mathbb{N}_{\geq 0} \rightarrow \mathcal{D}$ satisfying $\mathbf{d}_1 \leq_{D^w} \mathbf{d}_2$, we have $\Phi_{\kappa_1}(.,x_1^0,\mathbf{d}_1) \leq_{\mathcal{X}^w} \Phi_{\kappa_2}(.,x_2^0,\mathbf{d}_2)$.
\end{lemma}

\section{Controlled Invariants}
\label{sec:4}

We start by recalling the concept of controlled invariant~\cite{blanchini2008set}. In simple words, a controlled invariant set is a set for which there exists a controller such that if the state of the system is initialized in this set then its solutions remain there for all time.

\begin{definition}
 \label{def:contr_inv}
 Consider the system $\Sigma$ in (\ref{dis_sys}) and let $X \subseteq \mathcal{X}$, $U \subseteq \mathcal{U}$ and $D \subseteq \mathcal{D}$ be the constraints sets on the states, inputs and disturbances, respectively. The set $K \subseteq \mathcal{X}$ is said to be a {\it robust controlled invariant} for the system $\Sigma$ and constraint set $(X,U,D)$ if $K \subseteq X$ and there exists a controller $\kappa:X \rightarrow U$, with $\dom(\kappa)=K$ and such that for all $x_0 \in K$ and for any disturbance input $\mathbf{d}:\mathbb{N}_{\geq 0} \rightarrow D$ the solution of the closed loop system $\Phi_\kappa(.,x_0,\mathbf{d}):\mathbb{N}_{\geq 0} \rightarrow \mathcal{X}$ satisfies $\Phi_\kappa(k,x_0,\mathbf{d}) \in K$ for all $k\in \mathbb{N}_{\geq 0}$ \footnote{The condition $\Phi_\kappa(k,x_0,\mathbf{d}) \in K$ for all $k\in \mathbb{N}_{\geq 0}$ can be equivalently replaced by the following condition: $\Phi_\kappa(k,x_0,\mathbf{d}) \in X$ for all $k\in \mathbb{N}_{\geq 0}$.}. In this case, $\kappa$ is said to be {\it an invariance controller} for the system $\Sigma$ and constraint set $(X,U,D)$.
\end{definition}

While the characterization of controlled invariants in Definition~\ref{def:contr_inv} is the most commonly used in the literature~\cite{blanchini2008set,rungger2017computing}, the equivalent characterization below has also been used in the literature~\cite{rakovic2010parameterized}.

\begin{proposition}
\label{prop:charact}
Consider the system $\Sigma$ in (\ref{dis_sys}) and let $X \subseteq \mathcal{X}$, $U \subseteq \mathcal{U}$ and $D \subseteq \mathcal{D}$ be the constraints sets on the states, inputs and disturbances, respectively. The set $K \subseteq \mathcal{X}$ is a robust controlled invariant for the system $\Sigma$ and constraint set $(X,U,D)$ if and only if $K \subseteq X$ and the following holds:
\begin{equation}
\label{eqn:prop_charact}
    \forall x \in K,~ \exists u\in U~ \text{ s.t }~ f(x,u,D) \subseteq K.
\end{equation}
\end{proposition}

From Proposition~\ref{prop:charact}, one can readily see that the robust controlled invariance property is closed under union. Hence, there exists a unique robust controlled invariant that is maximal, in the sense that it contains all the robust controlled invariants. 
\begin{definition}
\label{def:contr_invar}
Consider the system $\Sigma$ in (\ref{dis_sys}) and let $X \subseteq \mathcal{X}$, $U \subseteq \mathcal{U}$ and $D \subseteq \mathcal{D}$ be the constraints sets on the states, inputs and disturbances, respectively. The set $K \subseteq \mathcal{X}$ is the \emph{maximal} robust controlled invariant set for the system $\Sigma$ and constraint set $(X,U,D)$ if:
\begin{itemize}
    \item $K \subseteq \mathcal{X}$ is a robust controlled invariant for the system $\Sigma$ and constraint set $(X,U,D)$;
    \item $K$ contains any robust controlled invariant for the system $\Sigma$ and constraint set $(X,U,D)$.
\end{itemize}
In this case, any invariance controller $\kappa:X \rightarrow U$ satisfying $\dom(\kappa)=K$ is said to be a maximal invariance controller for the system $\Sigma$ and constraint set $(X,U,D)$. 
\end{definition}

% We have the following auxiliary lemma characterizing the maximal invariance controller. The result follows immediately from the definitions above.

% \begin{lemma}
% \label{lem:cont_comparison}
% Consider the system $\Sigma$ in (\ref{dis_sys}) and let $X \subseteq \mathcal{X}$, $U \subseteq \mathcal{U}$ and $D \subseteq \mathcal{D}$ be the constraints sets on the states, inputs and disturbances, respectively. If $\kappa:X \rightarrow U$ is an invariance controller for the system $\Sigma$ and constraint set $(X,U,D)$, and $\kappa_m:X \rightarrow U$ is a maximal invariance controller for the system $\Sigma$ and constraint set $(X,U,D)$, then $\dom(\kappa) \subseteq \dom(\kappa_m)$.
% \end{lemma}

The following auxiliary result characterizes the effect of enlarging the set of control inputs and shrinking the set of disturbance inputs on the robust controlled invariance problem. The proof of this result is straightforward and thus omitted.

\begin{lemma}
\label{lem:inv_cont_dist}
Consider the system $\Sigma$ in (\ref{dis_sys}) and let $X \subseteq \mathcal{X}$, $U_1,U_2 \subseteq \mathcal{U}$ and $D_1,D_2 \subseteq \mathcal{D}$ be constraints sets on the states, inputs and disturbances, respectively, satisfying $U_1 \subseteq U_2$ and $D_2 \subseteq D_1$. If $K$ is a robust controlled invariant for the system $\Sigma$ and constraint set $(X,U_1,D_1)$ then $K$ is a robust controlled invariant for the system $\Sigma$ and constraint set $(X,U_2,D_2)$.
\end{lemma}
\section{Set-based characterization of Controlled Invariants}
\label{sec:5}
First, we have the following general characterization of the topological structure of controlled invariants for nonlinear systems under more regularity on the dynamics of the system.
\begin{proposition}
\label{prop:closedness}
Consider the system $\Sigma$ in (\ref{dis_sys}) and let $X \subseteq \mathcal{X}$, $U \subseteq \mathcal{U}$ and $D \subseteq \mathcal{D}$ be constraints sets on the states, inputs and disturbances, respectively. Suppose that the map $f:\mathcal{X}\times \mathcal{U}\times \mathcal{D}$ describing the system $\Sigma$ is lower semicontinuous on its first argument and the set of control inputs $U$ and disturbance inputs $D$ are compact. The following properties hold:
\begin{itemize}
    \item[(i)] If the set $K \subseteq X$ is a robust controlled invariant for the system $\Sigma$ and constraint set $(X,U,D)$, then the set $\cl(K)$ is a robust controlled invariant for the system $\Sigma$ and constraint set $(X,U,D)$;
    \item[(ii)] If the set $K \subseteq X$ is the maximal robust controlled invariant for the system $\Sigma$ and constraint set $(X,U,D)$, then the set $K$ is closed.
\end{itemize}
\end{proposition}

In the following, we provide different characterizations of robust controlled invariants when dealing with monotone dynamical systems and lower-closed safety specifications (i.e. lower closed set of constraints on the state-space $X$) .

\begin{theorem}
\label{thm1}
Consider the system $\Sigma$ in (\ref{dis_sys}) and let $X \subseteq \mathcal{X}$, $U \subseteq \mathcal{U}$ and $D \subseteq \mathcal{D}$ be the constraints sets on the states, inputs and disturbances, respectively, where the set $X$ is lower closed. The following properties hold:
\begin{itemize}
    \item[(i)] If the system $\Sigma$ is SM and if a set $K$ is a robust controlled invariant of the system $\Sigma$ and constraint set $(X,U,D)$, then its lower closure is also a robust controlled invariant for the system $\Sigma$ and constraint set $(X,U,D)$;
    
    \item[(ii)] If the system $\Sigma$ is SM then the maximal robust controlled invariant $K$ for the system $\Sigma$ and constraint set $(X,U,D)$ is lower closed;

    \item[(iii)] If the system $\Sigma$ is DSM and the set of disturbance inputs $D$ is closed and bounded above then the maximal robust controlled invariant for the system $\Sigma$ and constraint set $(X,U,D)$ is the maximal robust controlled invariant for the system $\Sigma$ and the constraint set $(X,U,D_{\max})$, where $D_{\max}=\max(D)$;

    \item[(iv)] If the system $\Sigma$ is CSM, the set of control inputs $U$ is closed and bounded below then the maximal robust controlled invariant for the system $\Sigma$ and constraint set $(X,U,D)$ is the maximal robust controlled invariant for the system $\Sigma$ and the and constraint set $(X,U_{\min},D)$, where $U_{\min}=\min(U)$;
    
    \item[(v)] If the system $\Sigma$ is CDSM, the set of control inputs $U$ is closed and bounded below and the set of disturbance inputs $D$ is closed and bounded above, then the maximal robust controlled invariant for the system $\Sigma$ and constraint set $(X,U,D)$ is the maximal robust controlled invariant for the system $\Sigma$ and constraint set $(X,U_{\min},D_{\max})$, where $U_{\min}=\min(U)$ and $D_{\max}=\max(D)$.
\end{itemize}
\end{theorem}

The result in (ii) states that for SM systems, the maximal robust controlled invariant can be characterized using only its maximal values (in the sense of the partial order $\leq_{\mathcal{X}}$). The result in (iii) states that to compute the maximal robust controlled invariant for DSM systems, it is sufficient to use maximal disturbance inputs. Finally, the result in (v) states that to compute the maximal robust controlled invariant for CDSM systems, it is sufficient to use maximal disturbance inputs and minimal control inputs. We also have the following characterizations of controlled invariants for the considered classes of systems.

\begin{proposition}
\label{prop:charac2}
Consider the system $\Sigma$ in (\ref{dis_sys}) and let $X \subseteq \mathcal{X}$, $U \subseteq \mathcal{U}$ and $D \subseteq \mathcal{D}$ be the constraints sets on the states, inputs and disturbances, respectively, where the set $X$ is lower closed. Consider a closed and lower closed set $K \subseteq X$. The following properties hold:
\begin{itemize}
    \item[(i)] If the system $\Sigma$ is SM then the set $K$ is a robust controlled invariant of the system $\Sigma$ and constraint set $(X,U,D)$, if and only if the following holds:
    \begin{equation}
    \label{eqn:prop_charact1}
    \forall x \in \max(K),~ \exists u\in U~ \text{ s.t }~ f(x,u,D) \subseteq K,
    \end{equation}

    \item[(ii)] If the system $\Sigma$ is DSM and the set of disturbance inputs $D$ is closed and bounded above then the set $K$ is a robust controlled invariant of the system $\Sigma$ and constraint set $(X,U,D)$, if and only if the following holds:
    \begin{equation}
    \label{eqn:prop_charact2}
    \forall x \in \max(K),~ \exists u\in U~ \text{ s.t }~ f(x,u,D_{\max}) \subseteq K,
    \end{equation}
    where $D_{\max}=\max(D)$;
    
    \item[(iii)] If the system $\Sigma$ is CSM and the set of control inputs $U$ is closed and bounded below then the set $K$ is a robust controlled invariant of the system $\Sigma$ and constraint set $(X,U,D)$, if and only if the following holds:
    \begin{equation}
    \label{eqn:prop_charact3}
    \forall x \in \max(K),~ \exists u\in U_{\min}~ \text{ s.t }~ f(x,u,D) \subseteq K,
    \end{equation}
    where $U_{\min}=\min(U)$;

    \item[(iv)] If the system $\Sigma$ is CDSM, the set of control inputs $U$ is closed and bounded below and the set of disturbance inputs $D$ is closed and bounded above then the set $K$ is a robust controlled invariant of the system $\Sigma$ and constraint set $(X,U,D)$, if and only if the following holds:
    \begin{equation}
    \label{eqn:prop_charact4}
    \forall x \in \max(K),~ \exists u\in U_{\min}~ \text{ s.t }~ f(x,u,D_{\max}) \subseteq K,
    \end{equation}
    where $U_{\min}=\min(U)$ and $D_{\max}=\max(D)$.
\end{itemize}
\end{proposition}

%Intuitively, Proposition~\ref{prop:charac2} can be interpreted as follows: the result in (i) ((ii), (iii) and (iv), respectively) states that the robust maximal controlled invariant for the system and the constraint set is the maximal lower-closed subset of $X$ for which condition (\ref{eqn:prop_charact1}), ((\ref{eqn:prop_charact2}), (\ref{eqn:prop_charact3}) and (\ref{eqn:prop_charact4}) respectively) is satisfied.

Proposition~\ref{prop:charac2} is critical from a computational point of view, when the objective is to check whether a lower-closed set is a robust controlled invariant. Indeed, while the invariance condition needs to be checked for all the elements $x \in K$, $u \in U$ and $d \in D$ for general nonlinear systems (see equation (\ref{eqn:prop_charact})), it has to be checked only for the elements:
\begin{itemize}
    \item $x \in \max(K)$, $u \in U$ and $d \in D$ for SM systems;
    \item $x \in \max(K)$, $u \in U$ and $d \in D_{\max}$ for DSM systems;
    \item $x \in \max(K)$, $u \in U_{\min}$ and $d \in D$ for CSM systems;
    \item $x \in \max(K)$, $u \in U_{\min}$ and $d \in D_{\max}$ for CDSM systems.
\end{itemize}
Moreover, one can also see that this property is very useful in practice when $\max(K)$, $D_{\max}$ and $U_{\min}$ are finite while $K$, $D$ and $U$ are infinite.

\section{Trajectory-based characterizations of controlled invariants}
\label{sec:6}

\begin{figure}[!t]
	\begin{center}
		\includegraphics[scale=0.6]{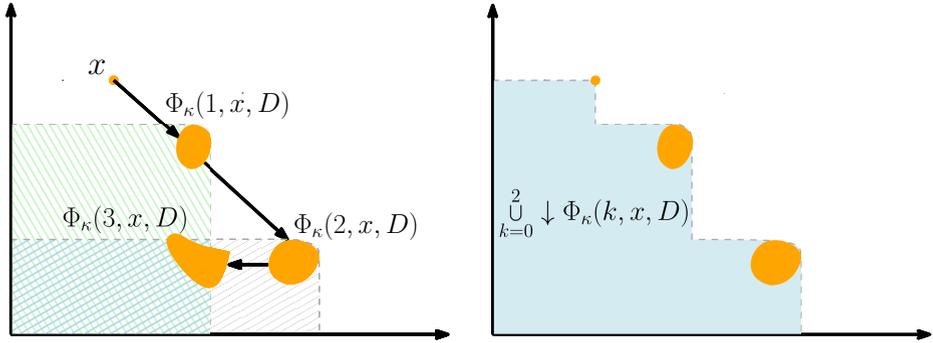}\\
	\end{center}
	\caption{Illustration of the concept of closed-loop feasibility. Left: the 3-step reachable set from the initial condition $x$ for a system $\Sigma$. Note that $\Phi_{\kappa}(3,x,D) \subseteq \downarrow \Phi_{\kappa}(1,x,D) \bigcup \downarrow \Phi_{\kappa}(2,x,D)$. Right: The set $\downarrow \bigcup\limits_{0 \leq k \leq 2} \Phi_{\kappa}(k,x_0,D)$ is a robust controlled invariant of the system $\Sigma$ in view of Proposition~\ref{prop:feas_inv}.}
	\label{fig:feasibility}	
\end{figure}

In this section we provide trajectory-based characterizations of controlled invariants. We start by introducing the concept of lower feasibiliy.

\begin{definition}
\label{def:feas}
Consider the system $\Sigma$ in (\ref{dis_sys}) and let $X \subseteq \mathcal{X}$, $U \subseteq \mathcal{U}$ and $D \subseteq \mathcal{D}$ be the constraints sets on the states, inputs and disturbances, respectively. A point $x_0 \in X$ is said to be {\it closed-loop feasible} w.r.t the constraint set $(X,U,D)$ if there exists a controller $\kappa:X \rightarrow U$ and $N \in \mathbb{N}_{>0}$ such that
\begin{equation}
\label{eqn:feas11}
    \Phi_{\kappa}(k,x_0,D) \subseteq X, \quad \forall~ k \in \{0,1,\ldots, N-1\},
\end{equation}
and 
\begin{equation}
\label{eqn:feas12}
 \Phi_{\kappa}(N,x_0,D) \subseteq \downarrow \bigcup\limits_{0 \leq k \leq N-1} \Phi_{\kappa}(k,x_0,D),
\end{equation}
Similarly, a point $x_0 \in X$ is said to be {\it open-loop feasible} w.r.t the constraint set $(X,U,D)$ if there exists an input trajectory $\mathbf{u}:\mathbb{N}_{\geq 0} \rightarrow \mathcal{U}$ and $N \in \mathbb{N}_{>0}$ such that 
\begin{equation}
\label{eqn:feas1o}
    \Phi(k,x_0,\mathbf{u},D) \subseteq X, \quad \forall~ k \in \{1,2,\ldots,N-1\}
\end{equation}
and 
\begin{equation}
\label{eqn:feas2o}
 \Phi(N,x_0,\mathbf{u},D) \subseteq \downarrow \bigcup\limits_{0 \leq k \leq N-1} \Phi(k,x_0,\mathbf{u},D).
\end{equation}
\end{definition}

% \begin{definition}
% Consider the system $\Sigma$ in (\ref{dis_sys}) and let $X \subseteq \mathcal{X}$, $U \subseteq \mathcal{U}$ and $D \subseteq \mathcal{D}$ be the constraints sets on the states, inputs and disturbances, respectively. An element $x_0 \in X$ is said to be feasible w.r.t the constraint set $(X,U,D)$ if there exist an $N \in \mathbb{N}_{>0}$ and a sequence of inputs $\sigma_u=u_0u_1\ldots u_{N-1}$ such that
% \begin{equation}
% \label{eqn:feas1}
%     \Phi_{\sigma_u}(k,x_0,D) \subseteq X, \quad \forall~ 0 \leq k \leq N-1
% \end{equation}
% and 
% \begin{equation}
% \label{eqn:feas2}
%  \Phi_{\sigma_u}(N,x_0,D) \subseteq \downarrow \bigcup\limits_{0 \leq k \leq N-1} \Phi_{\sigma_u}(k,x_0,D)
% \end{equation}
% \end{definition}

An illustration of the concept of closed-loop feasibility is presented in Figure~\ref{fig:feasibility} (Left). The following proposition shows the usefulness of feasible points to compute controlled invariants for SM systems. Conclusion (i) of this proposition is illustrated in Figure~\ref{fig:feasibility} (right).

\begin{proposition}
\label{prop:feas_inv}
Consider the system $\Sigma$ in (\ref{dis_sys}) and let $X \subseteq \mathcal{X}$, $U \subseteq \mathcal{U}$ and $D \subseteq \mathcal{D}$ be the constraints sets on the states, inputs and disturbances, respectively, where the set $X$ is lower closed. If the system $\Sigma$ is SM, then the following holds:
\begin{itemize}
    \item[(i)] If $x_0$ is closed loop feasible w.r.t the constraint set $(X,U,D)$ then there exists a controller $\kappa:X \rightarrow U$ and $N \in \mathbb{N}_{>0}$ such that the set  \begin{equation} K=\downarrow \bigcup\limits_{0 \leq k \leq N-1} \Phi_{\kappa}(k,x_0,D)    \end{equation} is a robust controlled invariant for the system $\Sigma$ and constraint set $(X,U,D)$.

    \item[(ii)] If $x_0 \in X$ is open-loop feasible w.r.t the constraint set $(X,U,D)$, then there exists an input trajectory $\mathbf{u}:\mathbb{N}_{\geq 0} \rightarrow \mathcal{U}$ and $N \in \mathbb{N}_{>0}$ such that the set  \begin{equation} K=\downarrow \bigcup\limits_{0 \leq k \leq N-1} \Phi(k,x_0,\mathbf{u},D)    \end{equation} is a robust controlled invariant for the system $\Sigma$ and constraint set $(X,U,D)$.
\end{itemize}

\end{proposition}

In the following, we characterize open-loop feasibility for DSM systems. Before providing the result, we give the following auxiliary lemma.

\begin{lemma}
\label{lem:DSM_feas}
Consider the system $\Sigma$ in (\ref{dis_sys}). If the system $\Sigma$ is DSM and the set of disturbance inputs $D$ is closed and bounded above then for any point $x_0 \in X$ and any input trajectory $\mathbf{u}:\mathbb{N}_{\geq 0} \rightarrow \mathcal{U}$, we have $\Phi(k,x_0,\mathbf{u},D) \subseteq \downarrow \Phi(k,x_0,\mathbf{u},D_{\max})$, for all $k \in \mathbb{N}_{\geq 0}$.
\end{lemma}

\begin{proposition}
\label{prop:feas}
Consider the system $\Sigma$ in (\ref{dis_sys}) and let $X \subseteq \mathcal{X}$, $U \subseteq \mathcal{U}$ and $D \subseteq \mathcal{D}$ be the constraints sets on the states, inputs and disturbances, respectively. If the system $\Sigma$ is DSM, the set of states $X$ is lower closed and the set of disturbance inputs $D$ is closed and bounded above, then a point $x_0 \in X$ is open-loop feasible w.r.t the constraint set $(X,U,D_{\max})$ if and only if it is open-loop feasible w.r.t the constraint set $(X,U,D)$, where $D_{\max}=\max(D)$.
\end{proposition}

We also have the following characterization of open-loop feasibility for a particular class of CSM systems.

\begin{proposition}
\label{prof:feas_CSM}
Consider the system $\Sigma$ in (\ref{dis_sys}) and let $X \subseteq \mathcal{X}$, $U \subseteq \mathcal{U}$ and $D \subseteq \mathcal{D}$ be the constraints sets on the states, inputs and disturbances, respectively, where $X$ is lower closed. If the system $\Sigma$ is CDSM, the set of states $X$ is lower closed, the set of control inputs $U$ is closed and bounded below, and for all $\varepsilon \in \mathbb{R}^n_{\geq 0}$, for all $x_1,x_2 \in \mathcal{X}$ and for all $u \in U$, following condition is satisfied:
\begin{equation}
    \label{eqn:feas_CSM_C}
    x_1 \geq x_2 +\varepsilon \implies \mathcal{B}_{\varepsilon}(f(x_2,u,D))\subseteq \downarrow f(x_1,u,D)
\end{equation}
then a point $x_0 \in X$ is open-loop feasible w.r.t the constraint set $(X,U,D)$ if and only if it is open-loop feasible w.r.t the constraint set $(X,U_{\min},D)$, where $U_{\min}=\min(U)$.
\end{proposition}

In the following result, we provide a comparison between the closed-loop and open-loop feasibility.

\begin{proposition}
\label{pro:feas_2}
Consider the system $\Sigma$ in (\ref{dis_sys}) and let $X \subseteq \mathcal{X}$, $U \subseteq \mathcal{U}$ and $D \subseteq \mathcal{D}$ be the constraints sets on the states, inputs and disturbances, respectively. If a point $x_0 \in X$ is open-loop feasible w.r.t the constraint set $(X,U,D)$ then it is closed-loop feasible w.r.t the constraint set $(X,U,D)$. Moreover, the converse holds if one of the following conditions is satisfied:
\begin{itemize}
    \item The system $\Sigma$ is disturbance free, i.e, $D=\{\overline{d}\}$.
    \item The system $\Sigma$ is DSM and $\card(D_{\max})=1$.
\end{itemize}
\end{proposition}

Moreover, we have the following result, characterizing a special case of open-loop feasibility for the particular class of monotone systems with Lipschitz dynamics.

\begin{theorem}
\label{thm:stric_feas}
Consider the SM system $\Sigma$ in (\ref{dis_sys}) and let $X \subseteq \mathcal{X}$, $U \subseteq \mathcal{U}$ and $D \subseteq \mathcal{D}$ be the constraints sets on the states, inputs and disturbances, respectively. Assume that the map $f: \mathcal{X} \times \mathcal{U} \times \mathcal{D} \rightarrow \mathcal{X}$ defining the system $\Sigma$ is continuous on its first and third arguments, and the sets of control inputs $U$ and disturbance inputs $D$ are compact. Consider $x_0 \in \mathcal{X}$. If the following conditions are satisfied:
\begin{itemize}
    \item[(i)] $x_0$ is open-loop feasible w.r.t the constraint set $(X,U,D)$ and there exists $\mathbf{u}:\mathbb{N}_{\geq 0} \rightarrow \mathcal{U}$, $N \in \mathbb{N}_{>0}$ and $\varepsilon_N$ such that 
\begin{equation}
\label{eqn:thm}
 \mathcal{B}_{\varepsilon_{N}}(\Phi_{\kappa}(N,x_0,\mathbf{u},D)) \subseteq \downarrow \bigcup\limits_{0 \leq k \leq N-1} \Phi_{\kappa}(k,x_0,\mathbf{u},D).
\end{equation}
    \item[(ii)] there exists $\gamma >0$ such that $ \mathcal{B}_{\gamma}(\Phi(k,x_0,\mathbf{u},D)) \subseteq X, \quad \forall~ k \in \{1,2,\ldots,N-1\}$
\end{itemize}
 then there exists $\beta>0$ such that for any $x_1 \in \{\uparrow x_0\} \cap \mathcal{B}_{\beta}(x_0)$, $x_1$ is open-loop feasible w.r.t the constraint set $(X,U,D)$. Moreover, when the map $f$ is $L$-Lipschitz on $\mathcal{X}$ on its first argument, then one can explicitly determine the value of $\beta$ as a function of the parameters $\varepsilon_N$ and $\gamma$. 
\end{theorem}

We have analyzed in Theorem~\ref{thm1} the structural properties of the maximal robust controlled invariant set for monotone systems and lower-closed safety specifications. In the following, and under more regularity assumptions on the system, we provide a characterization of the trajectories initiated at the boundaries of the maximal robust controlled invariant.

\begin{proposition}
\label{prop:charac_boundaries}
Consider the system $\Sigma$ in (\ref{dis_sys}) and let $X \subseteq \mathcal{X}$, $U \subseteq \mathcal{U}$ and $D \subseteq \mathcal{D}$ be the constraints sets on the states, inputs and disturbances, respectively, where the set $X$ is lower closed. Let $\kappa$ be a robust maximal invariance controller for the system $\Sigma$ and the constraint set $(X,U,D)$. If  the system $\Sigma$ is SM, the map $f: \mathcal{X} \times \mathcal{U} \times \mathcal{D} \rightarrow \mathcal{X}$ defining the system $\Sigma$ is continuous on its first and third arguments, and the sets of control inputs $U$ and disturbance inputs $D$ are compact, then for all $x_0 \in \partial(\dom(\kappa))$ one of the following scenarios hold:
\begin{itemize}
    \item[(i)] $\Phi_{\kappa}(k,x_0,D) \cap \partial(\dom(\kappa)) \neq \emptyset$ for all $k \in \mathbb{N}_{\geq 0}$;
    \item[(ii)] There exists $p \in \mathbb{N}_{> 0}$ such that 
    \begin{itemize}
        \item $\Phi_{\kappa}(p-1,x_0,D) \cap (\partial(\dom(\kappa))\setminus \partial X)= \emptyset$
        \item $\Phi_{\kappa}(p-1,x_0,D) \cap \partial X \neq \emptyset$
        \item $\Phi_{\kappa}(p,x_0,D) \subseteq \Int(\dom(\kappa))$.
    \end{itemize}
\end{itemize}
\end{proposition}

\begin{figure}[!t]
	\begin{center}
		\includegraphics[scale=0.7]{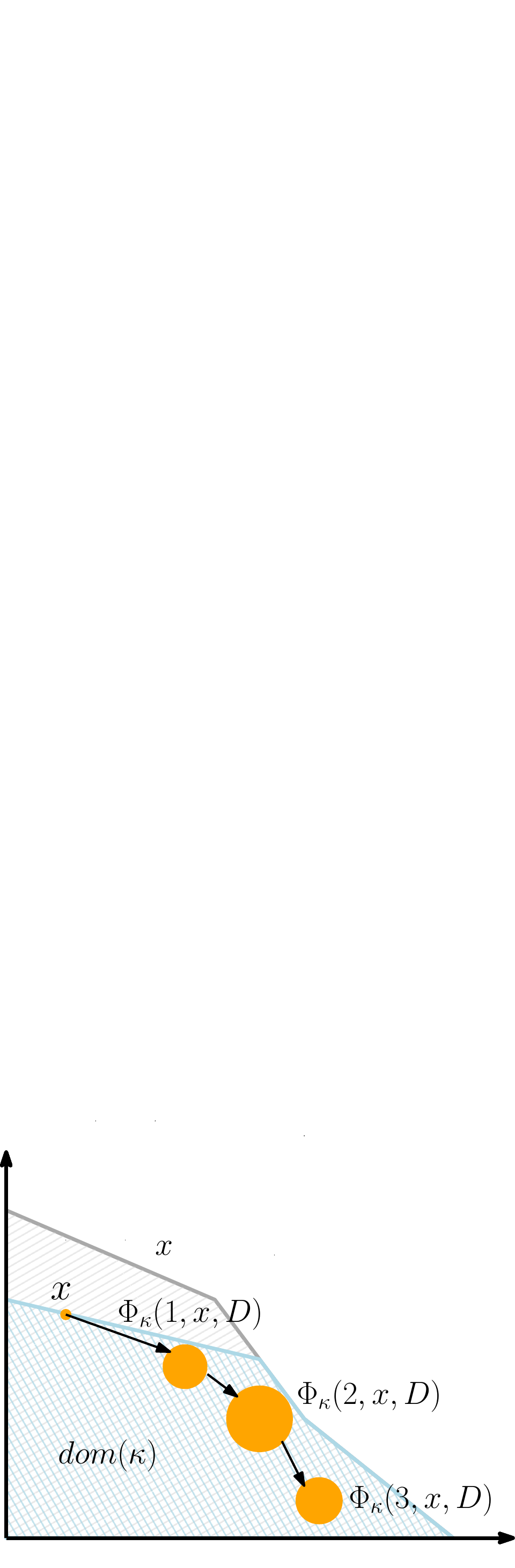}\\
	\end{center}
	\caption{An illustration of scenario (ii) in Proposition~\ref{prop:charac_boundaries} where we have a lower closed set $X$ (in gray) and the corresponding maximal robust controlled invariant set $K$ (in blue). For an initial condition $x \in \partial K$, the reachable set $\Phi_{\kappa}(.,x,D):\mathbb{N}_{\geq 0} \rightarrow X$ has an non-empty intersection with the boundaries of the set $\partial X$, before reaching the interior of the set $K$.}
	\label{fig:boundaries}	
\end{figure}

The result of Proposition~\ref{prop:charac_boundaries} is illustrated in Figure~\ref{fig:boundaries}. Intuitively, the result of Proposition~\ref{prop:charac_boundaries} states that for any initial condition $x$ in the boundary of the maximal robust controlled invariant $K$, either the reachable set $\Phi_{\kappa}(.,x,D):\mathbb{N}_{\geq 0} \rightarrow X$ will always have an non-empty intersection with the boundaries, which corresponds to case (i), or it will have an non-empty intersection with the boundaries of the set $X$ (i.e $\partial X$) before reaching the interior of the set $K$, which corresponds to case (ii).

\section{Verification and computation of controlled invariants}
\label{sec:7}

\subsection{Verification of controlled invariants}

The result of Proposition~\ref{prop:charac2} shows that robust controlled invariants for monotone dynamical systems enjoy useful properties, in the sense that the verification of whether a set is controlled invariant or no boils down to checking only the properties of its maximal elements. Based on this intuition, in this section, we present an algorithm to the verification of robust controlled invariants. In the following we present the main algorithm for the class of SM systems. Then, we explain how the algorithm can be improved for CSM and DSM systems.

Algorithm~\ref{algo1} works as follows: it explores all the elements $x$ of the set $\max(K)$. If for eaxh element $x \in \max (K)$, we have the existence of $u \in U$ such that $f(x,u,D) \subseteq K$, then, in view of Proposition~\ref{prop:charac2}, the set $K$ is a robust controlled invariant for the system $\Sigma$ and constraint set $(X,U,D)$; otherwise, if there exists $x \in \max(K)$ such that for all $u \in U$, $f(x,u,D) \nsubseteq K$, then the set $K$ is not a controlled invariant. 

Moreover, other properties can be used to improve the proposed algorithm. Indeed, if the system $\Sigma$ is DSM, then the condition $f(x,u,D) \subseteq K$ can be replaced by the simpler condition $f(x,u,D_{\max}) \subseteq K$, where $D_{\max}=\max(D)$. Similarly, if the system $\Sigma$ is CSM, then only the set of inputs $U_{\min}=\min(U)$ is explored, instead of the whole set of control inputs $U$.

\begin{algorithm}[!t]
\caption{{\bf Verification of controlled invariants}}
\label{algo1}
%\begin{algorithmic}[1]
{\bf Input:} A SM system $\Sigma$ as in (\ref{dis_sys}), a constraint set $(X,U,D)$, where $X$ a lower closed set, and a lower closed set $K \subset X$\\ {\bf Output:} True, if $K$ is a robust controlled invariant for the system $\Sigma$ and constraint set $(X,U,D)$.\\
1\hspace*{0.48cm}{\bf begin},\\
2\hspace*{0.48cm}\hspace*{0.3cm} $a:=1$,\\
3\hspace*{0.48cm}\hspace*{0.3cm}{\bf for} $x \in \max(K)$\\ 
4\hspace*{0.48cm}\hspace*{0.4cm}\hspace*{0.3cm}{\bf if} there exists $u \in U$ 
such that $f(x,u,D) \subseteq K$ \textbf{then} \\
5\hspace*{0.48cm}\hspace*{0.4cm}\hspace*{0.3cm}\hspace*{0.3cm} $a=a\times 1$\\
6\hspace*{0.48cm}\hspace*{0.4cm}\hspace*{0.3cm}{\bf else} $a=a \times 0$.\\
7\hspace*{0.48cm}\hspace*{0.4cm}\hspace*{0.3cm}\textbf{end if}\\
8\hspace*{0.48cm}\hspace*{0.3cm}{\bf end for}\\
9\hspace*{0.3cm}{\bf return} True if $a==1$.
%\end{algorithmic}
\end{algorithm}

\subsection{Computation of controlled invariants}

As shown in Theorem~\ref{thm1} and Proposition~\ref{prop:feas_inv}, controlled invariants for monotone systems and lower closed safety specifications are lower closed and can be computed using feasible points. This property implies that the boundary of the maximal robust controlled invariant set has the structure of a Pareto front and can therefore be approximated arbitrarily close, by resorting to multidimensional binary search algorithms used in multi-objective optimization~\cite{legriel2010approximating,tendulkar2014mapping}\footnote{Similar approaches, based on the approximation of the boundaries of Pareto fronts has been explored for the computation of timing and safety contracts in~\cite{al2017stability,zonetti2019Dc}.}. Based on such approaches, in the following we present the main algorithm for the computation of robust controlled invariants for the class of SM systems. Then, we explain the parts that needs to be modified for the case of DSM and CSM systems.

For a given $x$, the command {\it open loop-feasible} in Algorithm~\ref{algo} checks if $x$ is open-loop feasible, i.e. if it satisfies (\ref{eqn:feas2o}) for some input trajectory $\mathbf{u}:\mathbb{N}\rightarrow U$. If this is the case, any point in the lower closure of the set $Z$ defined below is feasible, and there is no need to explore it:
\begin{equation}
\label{eqn:Z}
    Z= \bigcup\limits_{0 \leq k \leq N-1} \Phi(k,x_0,\mathbf{u},D).
\end{equation}
Hence, all the elements of the set $Z$ are stored in the set $\mathcal{F}_1$ representing the set of states belonging to the maximal robust controlled invariant. Similarly, we use the set $\mathcal{F}_2$ to store the elements that do not belongs to the maximal robust controlled invariant set. The command {\it leads to the unsafe set} $\mathcal{F}_2\cup \overline{X}$ in Algorithm~\ref{algo} checks if for all possible input trajectories $\mathbf{u}:\mathbb{N}\rightarrow U$ there exists $k \in \{1,2\ldots,N\}$ such that $\Phi(k,x_0,\mathbf{u},D) \cap \mathcal{F}_2\cup \overline{X} \neq \emptyset$. If $x$ leads to the unsafe set, then any state from the upper closure of the set $Y$ defined below will lead to the unsafe set, and there is no need to explore it:
\begin{equation}
\label{eqn:Y}
    Y= \bigcup\limits_{0 \leq k \leq N-1} \Phi(k,x_0,\mathbf{u},D)
\end{equation}
Algorithm~\ref{algo} consists of three parts. In the first part (lines $2-11$), the elements of the set $\max(X)$ are explored. In the second part (lines $12-18$), the elements of the set $\min(X)$\footnote{Let us mention that in general, the set $X$ may not be bounded from below. In this case the set $\min(X)$ can be replaced by any collection of open loop feasible points, and which can be computed before running the algorithm.} are explored. Finally, lines $22-29$ describe the main loop of the algorithm. We start by picking an element from the set of non-explored points and for which we did not decide yet if they are open-loop feasible, or leading to the unsafe set. The strategy to pick a new point to explore is adapted from~\cite{legriel2010approximating}. The algorithm stops when the Hausdorff distance between the sets $\mathcal{F}_1$ and $\mathcal{F}_2$ is smaller than the precision parameter $\varepsilon>0$. In this case, we get that the set  $K=X \cap \mathcal{F}_1$ is a controlled invariant for the system $\Sigma$ and the constraint set $(X,U,D)$, and moreover, we also have that $K \subseteq K^* \subseteq \mathcal{B}_{\varepsilon}(K)$, where $K^*$ is the maximal robust controlled invariant for the system $\Sigma$ and the constraint set $(X,U,D)$. This last statement follows directly from the construction of the sets $\mathcal{F}_1$ and $\mathcal{F}_2$ since:
\begin{itemize}
    \item any element of $\mathcal{F}_1$ belongs to the maximal robust controlled invariant and which is due to the fact that it is constructed based on feasible points;
    \item any elements of $\mathcal{F}_2$ leads to the unsafe set, and do not belong to the maximal robust controlled invariant.
\end{itemize}

Let us now explain how other different structural properties of the system allow to improve the proposed algorithm; these details were removed from the description of the algorithm to improve its readability.

\begin{itemize}
    \item DSM systems: For both {\it open loop feasible} and {\it leads to unsafe set} commands, and in view of Proposition~\ref{prop:feas}, the trajectories of the system are computed only with respect to the set of maximal disturbances $D_{\max}=\max(D)$ when the system is DSM;
    \item CSM systems: For both {\it open loop feasible} and {\it leads to unsafe set}, and in view of Proposition~\ref{prof:feas_CSM}, the trajectories of the system are computed only with respect to the set of minimal inputs $U_{\min}=\min(U)$ when the system is CSM and condition (\ref{eqn:feas_CSM_C}) is satisfied;
    \item $L$-Lipschitz systems: For the {\it open-loop feasible} command and if for some $x_0 \in X$ conditions (i) and (ii) of Theorem~\ref{thm:stric_feas} are satisfied, with some $\varepsilon_N,\gamma>0$, any point in the set $\{\uparrow x_0\} \cap \mathcal{B}_{\beta}(x_0)$ is feasible, and there is no need to explore it, where $\beta$ is given in the proof of Theorem~\ref{thm:stric_feas} as a function of $\beta_N,\gamma$ and the Lipschitz constant $L$.
\end{itemize}

\begin{algorithm}[!t]
\caption{{\bf Computation of controlled invariants}}
\label{algo}
{\bf Input:} A SM system $\Sigma$ as in (\ref{dis_sys}), a contraint set $(X,U,D)$, where $X$ a lower closed set, and a precision $\varepsilon >0$\\ {\bf Output:} A controlled invariant set $K \subseteq X$ .\\
1\hspace*{0.48cm}{\bf begin}, $\mathcal{F}_1=\mathcal{F}_2=\emptyset$\\
2\hspace*{0.48cm}\hspace*{0.3cm}{\bf for} $x \in \max(X)$\\ 
3\hspace*{0.48cm}\hspace*{0.4cm}\hspace*{0.3cm}{\bf if} $x$ is open-loop feasible \textbf{then} \\
4\hspace*{0.48cm}\hspace*{0.4cm}\hspace*{0.3cm}\hspace*{0.3cm}
$\mathcal{F}_1=\mathcal{F}_1\cup \downarrow Z$, with $Z$ from (\ref{eqn:Z})\\
5\hspace*{0.48cm}\hspace*{0.4cm}\hspace*{0.3cm}{\bf else if} $x$ leads to the unsafe set $\mathcal{F}_2\cup \overline{X}$\\
6\hspace*{0.48cm}\hspace*{0.4cm}\hspace*{0.3cm}\hspace*{0.3cm}
$\mathcal{F}_2=\mathcal{F}_2\cup \uparrow Y$, with $Y$ from (\ref{eqn:Y})\\
7\hspace*{0.48cm}\hspace*{0.4cm}\hspace*{0.3cm}\textbf{end if}\\
8\hspace*{0.48cm}\hspace*{0.3cm}{\bf end for}\\
9\hspace*{0.48cm}\hspace*{0.3cm}{\bf if} $X=\mathcal{F}_1$\\
10\hspace*{0.3cm}\hspace*{0.4cm}\hspace*{0.3cm}{\bf return} $K=X$\\
11\hspace*{0.3cm}\hspace*{0.3cm}{\bf end if}\\
12\hspace*{0.3cm}\hspace*{0.3cm}{\bf for} $x \in \min(X)$\\ 
13\hspace*{0.3cm}\hspace*{0.4cm}\hspace*{0.3cm}{\bf if} $x$ is open-loop feasible \textbf{then} \\
14\hspace*{0.3cm}\hspace*{0.4cm}\hspace*{0.3cm}\hspace*{0.3cm}
$\mathcal{F}_1=\mathcal{F}_1\cup \downarrow Z$, with $Z$ from (\ref{eqn:Z})\\
15\hspace*{0.3cm}\hspace*{0.4cm}\hspace*{0.3cm}{\bf else if} $x$ leads to the unsafe set $\mathcal{F}_2\cup \overline{X}$\\
16\hspace*{0.3cm}\hspace*{0.4cm}\hspace*{0.3cm}\hspace*{0.3cm}
$\mathcal{F}_2=\mathcal{F}_2\cup \uparrow Y$, with $Y$ from (\ref{eqn:Y})\\
17\hspace*{0.3cm}\hspace*{0.4cm}\hspace*{0.3cm}\textbf{end if}\\
18\hspace*{0.3cm}\hspace*{0.3cm}{\bf end for}\\
19\hspace*{0.3cm}\hspace*{0.3cm}{\bf if} $\min(X) \subseteq \mathcal{F}_2$\\
20\hspace*{0.3cm}\hspace*{0.4cm}\hspace*{0.3cm}{\bf return} $K=\emptyset$\\
21\hspace*{0.3cm}\hspace*{0.3cm}{\bf end if}\\
22\hspace*{0.3cm}\hspace*{0.3cm}{\bf while} $\mathbf{d}(\mathcal{F}_2,\mathcal{F}_1)> \varepsilon$\\ 
23\hspace*{0.3cm}\hspace*{0.4cm}\hspace*{0.3cm}Pick $x' \in (X \setminus \mathcal{F}_2) \cap (X \setminus \mathcal{F}_1) $ \\
24\hspace*{0.3cm}\hspace*{0.4cm}\hspace*{0.3cm}{\bf if} $x$ is open-loop feasible \textbf{then} \\
25\hspace*{0.3cm}\hspace*{0.4cm}\hspace*{0.3cm}\hspace*{0.3cm}
$\mathcal{F}_2=\mathcal{F}_2\cup \uparrow Y$, with $Y$ from (\ref{eqn:Y})\\
26\hspace*{0.3cm}\hspace*{0.4cm}\hspace*{0.3cm}{\bf else if} $x$ leads to the unsafe set $\mathcal{F}_2\cup \overline{X}$\\
27\hspace*{0.3cm}\hspace*{0.4cm}\hspace*{0.3cm}\hspace*{0.3cm}
$\mathcal{F}_1=\mathcal{F}_1\cup \downarrow Z$, with $Z$ from (\ref{eqn:Z})\\
28\hspace*{0.3cm}\hspace*{0.4cm}\hspace*{0.3cm}\textbf{end if}\\
29\hspace*{0.3cm}\hspace*{0.4cm}\textbf{end while}\\
30\hspace{0.3cm}{\bf return} $K=X \cap \mathcal{F}_1$ .
\end{algorithm}

\section{Numerical Examples}
\label{sec:8}

\subsection{Verification of robust controlled invariants}

Consider the two dimensional switched system $\Sigma$ described by
\begin{equation*}
		x(k+1)=\;
		\left\{
		\begin{array}{c c}
 		A_1x(k)+d(k) \;\text{ if }\; u=1  \\
		A_2x(k)+d(k) \;\text{ if }\; u=2
		\end{array}
		\right. 
\end{equation*}
with $x(k) \in \mathbb{R}_{\geq 0}^2$, $d(k) \in \mathcal{D} \in [0, 0.2] \times [0, 0.2]$ and $u(k) \in \{1,2\}$
$$A_1=\begin{pmatrix} 1.2 & 0.1 \\ 0.2 & 0.5 \end{pmatrix} \text{ and } A_2=\begin{pmatrix} 0.4 & 0.1 \\ 0.1 & 1.1 \end{pmatrix}$$
One can easily check that if the input $u$ of the switched system is fixed, the trajectory will grow unbounded. The objective here is to verify that the set $K=\mathbb{R}_{\geq 0} \cap\{ \downarrow a_1\} \cup \{\downarrow a_2\} \cup \{ \downarrow a_3\}$ is a robust controlled invariant, with $a_1=[50, 25]$, $a_2=[25, 50]$ and $a_3=[36, 31]$.

\begin{figure}[!t]
	\begin{center}
		\includegraphics[width=0.8\columnwidth]{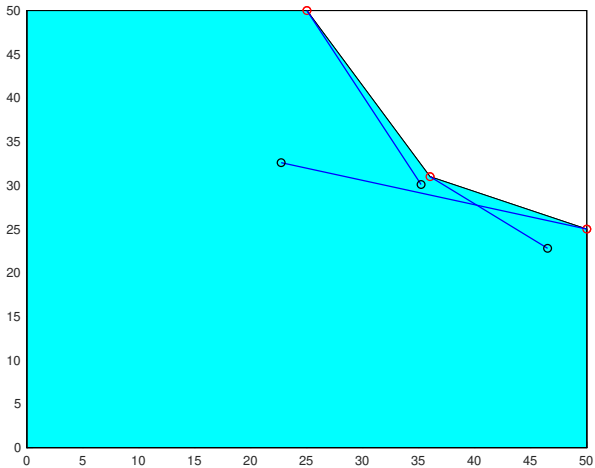}\\
	\end{center}
	\caption{\footnotesize{The light blue region represents set $K$. The three blue trajectories are initiated from the maximal points (in red) $a_1$, $a_2$ and $a_3$.}}
	\label{fig:verif}	
\end{figure}

Using Algorithm~\ref{algo1}, and by exploiting the fact that the system $\Sigma$ is DSM for the usual partial order on $\mathbb{R}^2_{\geq 0}$, we only explore the maximal elements of the set $K$, namely $a_1$, $a_2$ and $a_3$, while using the maximal disturbance input. Indeed, one can readily see in Figure~\ref{fig:verif} that $A_2a_1+\max(\mathcal{D})=[22.7, 35.2] \in K$, $A_1a_2+\max(\mathcal{D})=[35.2, 30.1] \in K$ and $A_1a_3+\max(\mathcal{D})=[35.2, 30.1] \in K$, which shows in view of Proposition~\ref{prop:charac2} that the set $K$ is a robust controlled invariant set for the system $\Sigma$ and constraint set $(X,U,D)$. {The computation time to implement Algorithm \ref{algo1} for this particular example is given by $0.18$ seconds. The resolution of the proposed verification problem using the classical safety fixed-point algorithm \cite{blanchini2008set,tabuada2009verification,rakovic2004computation} took almost $1$ second. The numerical implementations have been done in MATLAB and a computer with processor Apple M1 Max and Memory of 64 GB.}

\subsection{Computation of robust controlled invariants}

We consider a vehicle model moving along a straight road. The dynamics of the vehicle is adapted from~\cite{saoud2018contract} and described as:
%\small
\begin{equation}
\label{eqn:model}
m\dot{v}=\alpha(u,v)=\left\{
\begin{array}{l c r}
u-f_{0}-f_{2}v^2  &\text{ if }& v>0\\
\max(u-f_0,0)  &\text{ if }& v=0
\end{array}
\right.
\end{equation}
%\normalsize
where $m>0$ is the mass of the vehicle, $u$ is the net engine torque applied to the wheels, $v \geq 0$ represents the velocity of the vehicle and the term $f_0+f_2v^2$ includes the rolling resistance and aerodynamics. For this system, $u$ is the control input and satisfies $u \in [U_{\min},U_{\max}]$. Moreover, we include a lead vehicle whole velocity satisfies $d \in D$, is considered as a disturbance. The dynamics of the system is given by:
%\small
\begin{equation}
\label{eqn4}
\left\{
\begin{array}{r c l}
\dot{h}&=&d - v\\
m\dot{v}&=&\alpha(u,v).
\end{array}
\right.
\end{equation}
From this continuous-time system, we generate a discrete-time model using the sampling period $\tau=0.5 s$, while conserving the monotonicity property of the system.
%\normalsize
\begin{remark}
The system can be easily transformed to a CDSM system one by using the following change of coordinates: $d'=-d$ and $z=-h$.
\end{remark}% In this case, the dynamics of the system will be written as:
%\begin{equation}
%\left\{
%\begin{array}{r c l}
%\dot{h}&=&w + a\\
%m\dot{v}&=&\alpha(u,v).
%\end{array}
%\right.
%\end{equation}
%where $z \in [-v_{\max},0]=-W$.
%\end{remark}
%\medskip

\begin{table}
	\caption{{Vehicle and safety parameters}} 
	\centering
	\begin{tabular}{|c|c|c|}
		\hline 
		Parameter & Value & Unit \\ 
		\hline 
		$M$ & $1370$ & $Kg$ \\ 
		
		$f_0$ & $51.0709$ & $N$ \\ 
		
		$f_2$ & $0.4161$ &  $Ns^2/{m^2}$\\ 
		
		$U_{\min}$&  $-4031.9$ &   $mKg/{s^2}$ \\
		
		$U_{\max}$&  $2687.9$&   $mKg/{s^2}$ \\
		
		$d_{\min}$&  $10$&   $m$ \\
		
		$d'$&  $70$&   $m$  \\
		
		$v_{\max}$&  $15$&   $m/s$ \\
		\hline 
	\end{tabular}
	\label{table:parameters}
\end{table}

\begin{figure}[!t]
	\begin{center}
		\includegraphics[width=0.8\columnwidth]{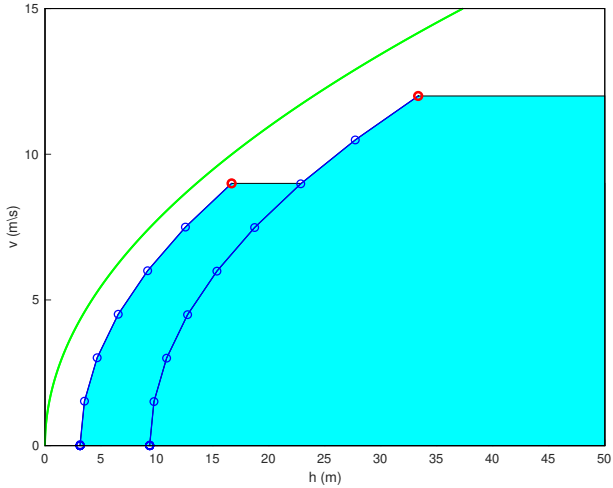}\\
	\end{center}
	\caption{\footnotesize{The light blue region represents the domain of the robust controlled invariant. The blue trajectories are initiated from two feasible points (in red) $x_1=[33.75; 13.5]$ and $x_2=[16.25; 9.75]$. The green curve represents the boundary of the maximal robust controlled invariant. The precision $\varepsilon$ chosen for Algorithm~\ref{algo} is $\varepsilon=1.5$.}}
	\label{fig:epsilon_a}	
\end{figure}

\begin{figure}[!t]
	\begin{center}
		\includegraphics[width=0.8\columnwidth]{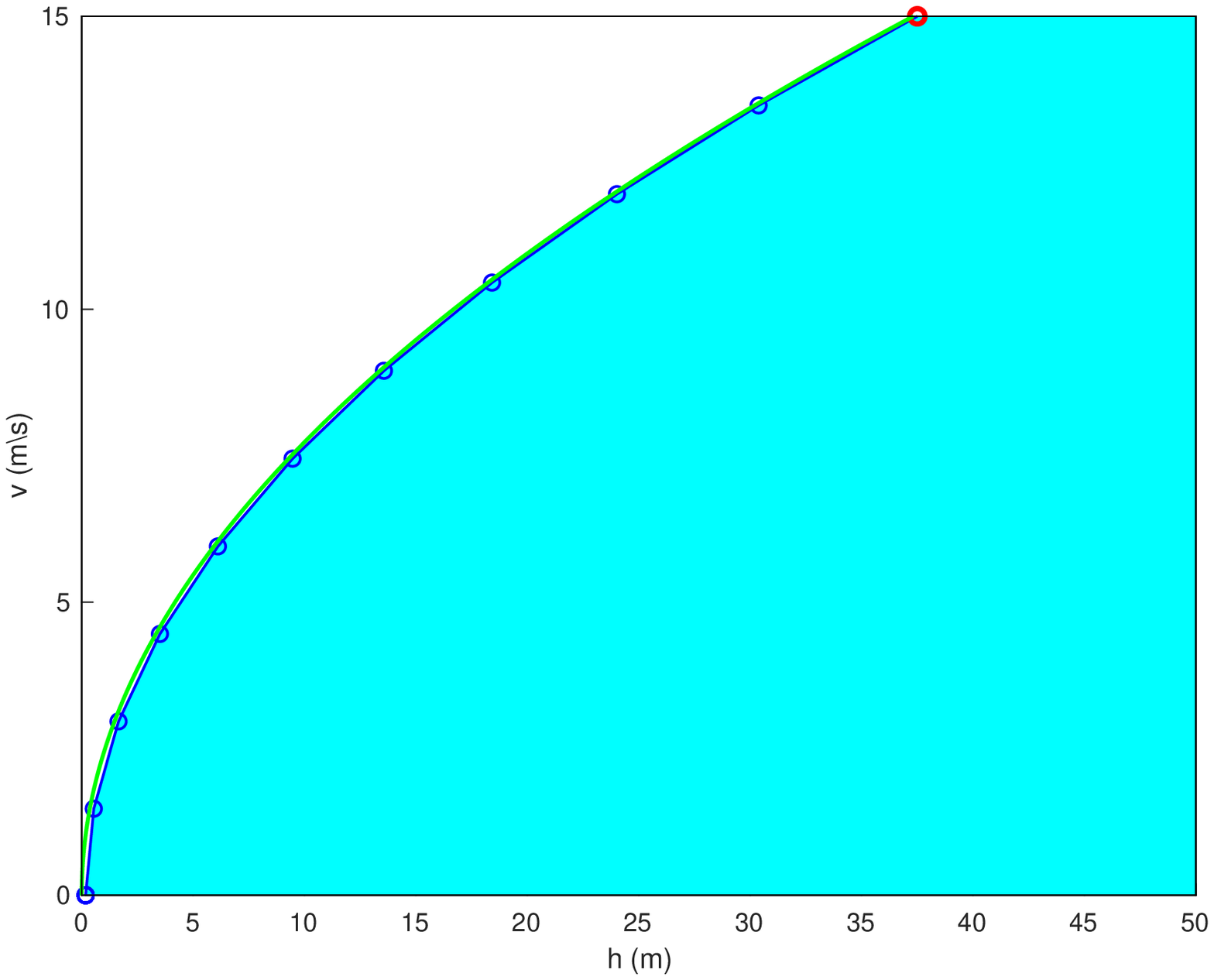}\\
	\end{center}
	\caption{\footnotesize{The light blue region represents the domain of the robust controlled invariant. The blue trajectory is initiated from the feasible point (in red) $x=[37.5; 15]$. The green curve represents the boundary of the maximal robust controlled invariant. The precision $\varepsilon$ chosen for Algorithm~\ref{algo} is $\varepsilon=0.01$.}}
	\label{fig:epsilon_b}	
\end{figure}

The objective is to compute a controlled invariant for the system in order to ensure that the velocity remains between $0$ and $v_{\max}$, and the relative distance between the leader and the follower remains larger than $0$, while assuming that the velocity of the leader $d$ belongs to the set $D=[0,v_{\max}]$. Moreover, since the constraint $v \geq 0$ is directly satisfied from the model description in~(\ref{eqn:model}), the constraint set is a lower closed set. For the computation of a robust controlled invariant, we use Algorithm~\ref{algo}. The parameters model are taken from~\cite{saoud2018contract} and are presented in Table~\ref{table:parameters}. Figures~\ref{fig:epsilon_a} and~\ref{fig:epsilon_b} represent the computed robust controlled invariant set for two different precisions $\varepsilon=1.5$ and $\varepsilon=0.01$. Moreover, we also present in green the boundary of the maximal robust controlled invariant set, which can be computed analytically for this problem, following the approach presented in~\cite{devonport2020data}. One can see in Figure \ref{fig:epsilon_b} that the domain of the obtained controller using a precision $\varepsilon=0.01$ is almost the same as the domain of the maximal safety controller.


\section{Conclusion}
In this paper, we have presented different characterizations of robust controlled invariants for discrete-time monotone dynamical systems, together with an algorithmic procedure to compute the invariants for the considered class of systems. An illustrative example is presented showing the merits of the proposed approach.  In future work, we will develop more general algorithms allowing to extend the approach from safety to other types of specifications, such as stability or more general properties described by signal temporal logic
formulas.

\begin{appendices}

\section{Proofs}

\textbf{\underline{Proof of Proposition~\ref{prop:characterizations_monotone}:}}

We only show (iv), the proofs of (i), (ii) and (iii) can be derived similarly. Let us start with the sufficient condition. Assume that $\Sigma$ is a CDSM system and consider $x',x \in \mathcal{X}$, $u',u \in \mathcal{U}$ and $d',d \in \mathcal{D}$ satisfying $x' \in (\downarrow x)$, $u' \in (\downarrow u)$ and $d' \in (\downarrow d)$. Using the fact that $\Sigma$ is a CDSM system, we have that $f(x',u', d')\leq_{\mathcal{X}} f(x,u,d)$, which in turn implies that $f(\downarrow x,\downarrow u,\downarrow d) \subseteq \downarrow f(x,u,d)$. Let us now show the necessary condition. Let $x_1,x_2 \in \mathcal{X}$, $u_1,u_2 \in \mathcal{U}$ and $d_1,d_2 \in \mathcal{D}$ with $x_1 \leq_\mathcal{X} x_2$, $u_1\leq_\mathcal{U} u_2$ and $d_1 \leq_{\mathcal{D}} d_2$ and let us show that $\Sigma$ is CDSM. We have that $x_1\in  (\downarrow x_2)$, $u_1 \in (\downarrow u_2)$ and $d_1 \in (\downarrow d_2)$. Then, from our assumption, we have that $f(x_1,u_1,d_1) \in f(\downarrow x_2,\downarrow u_2, \downarrow d_2) \subseteq  \downarrow f(x_2,u_2,d_2)$, which in turn implies that $f(x_1,u_1,d_1) \leq_{\mathcal{X}} f(x_2,u_2,d_2)$. Hence, $\Sigma$ is a CDSM system and (iv) holds.

\bigskip

\textbf{\underline{Proof of Lemma~\ref{lem:cont_mono}:}}

Consider $x_1^0,x_2^0 \in \mathcal{X}$, with $x_1^0 \leq_{\mathcal{X}} x_2^0$ and  $\mathbf{d}_1,\mathbf{d}_2:\mathbb{N}_{\geq 0} \rightarrow \mathcal{D}$ satisfying $\mathbf{d}_1 \leq_{D^w} \mathbf{d}_2$. To show the result we proceed by induction. First, we have that $\Phi_{\kappa_1}(0,x_1^0,\mathbf{d}_1)=x_1^0 \leq_{\mathcal{X}} x_2^0= \Phi_{\kappa_2}(0,x_2^0,\mathbf{d}_2)$. Now, consider $k \in \mathbb{N}_{\geq 0}$ and assume that $\Phi_{\kappa_1}(k,x_1^0,\mathbf{d}_1) \leq_{\mathcal{X}} \Phi_{\kappa_2}(k,x_2^0,\mathbf{d}_2)$. Since the system $\Sigma$ is CDSM and using (\ref{eqn:cont_mono}) and the fact that $\mathbf{d}_1 \leq_{D^w} \mathbf{d}_1$ we have that 
\small
\begin{align*}
    \Phi_{\kappa_1}(k+1,x_1^0,\mathbf{d}_1)&=f(\Phi_{\kappa_1}(k,x_1^0,\mathbf{d}_1),\kappa_1(\Phi_{\kappa_1}(k,x_1^0,\mathbf{d}_1)),\mathbf{d}_1(k))\\ &\leq_{\mathcal{X}} f(\Phi_{\kappa_2}(k,x_2^0,\mathbf{d}_2),\kappa_2(\Phi_{\kappa_2}(k,x_2^0,\mathbf{d}_2)),\mathbf{d}_2(k))\\ &= \Phi_{\kappa_2}(k+1,x_2^0,\mathbf{d}_2)
\end{align*}
\normalsize
Hence, $\Phi_{\kappa_1}(.,x_1^0,\mathbf{d}_1) \leq_{\mathcal{X}^w} \Phi_{\kappa_2}(.,x_2^0,\mathbf{d}_2)$.

\bigskip

\textbf{\underline{Proof of Proposition~\ref{prop:charact}:}}

\textbf{\underline{Sufficient condition:}} Consider a controller $\kappa:X \rightarrow U$ defined as $$\kappa(x):=\Single(\{u \in U \mid f(x,u,d) \in K \text{ for all } d \in D \})$$ Let us show that $\kappa$ is a robust invariance controller for the system $\Sigma$ and constraint set $(X,U,D)$. Consider $x_0 \in K$ and $\mathbf{d}:\mathbb{N}_{\geq 0} \rightarrow D$ and let us show by induction that $\Phi_\kappa(k,x_0,\mathbf{d}) \in K$ for all $k \in \mathbb{N}_{\geq 0}$. First, we have that $\Phi_\kappa(0,x_0,\mathbf{d})=x_0 \in K$. Now assume that $\Phi_\kappa(k,x_0,\mathbf{d}) \in K$ and let us show that $\Phi_\kappa(k+1,x_0,\mathbf{d}) \in K$. Since $\Phi_\kappa(k,x_0,\mathbf{d}) \in K$, we have the existence of $u=\kappa(\Phi_\kappa(k,x_0,\mathbf{d})) \in U$ such that for $\mathbf{d}(k)\in D$, $\Phi_\kappa(k+1,x_0,\mathbf{d})=f(\Phi_\kappa(k,x_0,\mathbf{d}),\kappa(\Phi_\kappa(k,x_0,\mathbf{d})),\mathbf{d}(k))\in K$. Hence, $\kappa$ is a robust invariance controller for the system $\Sigma$ and constraint set $(X,U,D)$.

\textbf{\underline{Necessary condition:}} Assume the existence of a controller $\kappa:X \rightarrow U$, with $\dom(\kappa)=K$ and such that for all $x_0 \in K$ and for any disturbance input $\mathbf{d}:\mathbb{N}_{\geq 0} \rightarrow D$ the solution of the closed loop system $\Phi_\kappa(.,x_0,\mathbf{d}):\mathbb{N}_{\geq 0} \rightarrow \mathcal{X}$ satisfies $\Phi_\kappa(k,x_0,\mathbf{d}) \in K$ for all $k\in \mathbb{N}_{\geq 0}$. Consider $x \in K$, we have the existence of $u=\kappa(x)\in U$ such that $f(x_0,\kappa(x),d)=\Phi_\kappa(1,x_0,\mathbf{d})\in K$ for all $d \in D$, where $\mathbf{d}:\mathbb{N}_{\geq 0} \rightarrow D$ is any disturbance input trajectory satisfying $\mathbf{d}(0)=d\in D$, which ends the proof.

\bigskip

\textbf{\underline{Proof of Proposition~\ref{prop:closedness}:}}

We provide a proof for each item separately.
\textbf{\underline{Proof of (i):}}
Let $K$ be a robust controlled invariant for the system $\Sigma$ and constraint set $(X,U,D)$ and let us show, by contradiction, that $\cl(K)$ is a robust controlled invariant for the system $\Sigma$ and constraint set $(X,U,D)$. Consider $x \in \cl(K)\setminus K$ and assume that for all $u \in U$ we have that $f(x,u,D) \cap \overline{\cl(K)} \neq \emptyset$. Consider $u\in U$, since the set $\overline{\cl(K)}$ is open, we have the existence of $\varepsilon_u >0$ and $y_{x,u} \in f(x,u,D)\cap \overline{\cl(K)}$ such that $\mathcal{B}_{\varepsilon_u}(y_{x,u}) \subseteq \overline{\cl(K)}$. Since, the set $D$ is compact and $f$ is lower semicontinuous on its first argument, one has that the set valued map $F:\mathcal{X}\times \mathcal{U} \rightrightarrows \mathcal{X}$ defined for $x \in \mathcal{X}$ and $u\in \mathcal{U}$ by $F(x,u):=f(x,u,D)$ is lower semicontinuous on its first argument. Hence, for $\varepsilon_u >0$ and $y_{x,u} \in F(x,u)$, we have the existence of $\eta_u>0$ such that for all $z \in \mathcal{B}_{\eta_u}(x)$, there exists $y_{z,u} \in F(z,u)$ satisfying $y_{z,u} \in \mathcal{B}_{\varepsilon_u}(y_{x,u}) \subseteq \overline{\cl(K)}$. Now consider $\eta=\min_{u \in U}\eta_u$. Since the set $U$ is compact, we have that $\eta>0$. Hence, it follows from above that for any $z \in \Int(K) \cap \mathcal{B}_{\eta}(x)$ and for any $u \in U$, we have the existence of $y_{z,u} \in F(z,u)$ satisfying $y_{z,u} \in \overline{\cl(K)}$, which contradicts the robust controlled invariance of the set $K$. Hence, the set $\cl(K)$ is a robust controlled invariant for the system $\Sigma$ and constraint set $(X,U,D)$.

\textbf{\underline{Proof of (ii):}} Let $K$ be the maximal robust controlled invariant for the system $\Sigma$ and constraint set $(X,U,D)$. From (i) it follows that $\cl(K)$ is a robust controlled invariant for the system $\Sigma$ and constraint set $(X,U,D)$. Hence, it follows from maximality of the set $K$ that $\cl(K)=K$.

\bigskip

\textbf{\underline{Proof of Theorem~\ref{thm1}:}}

We provide a proof for each item separately.

\textbf{\underline{proof of (i):}} To show the result, we use the characterization of robust controlled invariance from Proposition~\ref{prop:charact}. Let $K$ be a robust controlled invariant for the system $\Sigma$ and constraint set $(X,U,D)$. Consider the set $H =\downarrow K$ and let us show that the set $H$ is a robust controlled invariant for the system $\Sigma$ and constraint set $(X,U,D)$. Consider $x \in H=\downarrow K$, we have the existence of $x' \in K $ such that $x \leq_{\mathcal{X}} x'$. Since $x' \in K$, which is a robust controlled invariant, we have from Proposition~\ref{prop:charact} the existence of $u \in U$ such that $f(x',u,d) \in K$ for all $d \in D$. Using the fact that $\Sigma$ is a SM system we have that $f(x,u,d) \leq_{\mathcal{X}} f(x',u,d)$ for all $d \in D$. Hence, it follows that $f(x,u,d) \in \downarrow K =H$, for all $d \in D$, which implies that $H=\downarrow K$ is a robust controlled invariant for the system $\Sigma$ and constraint set $(X,U,D)$.

 \textbf{\underline{proof of (ii):}} To show the result, we use the characterization of robust controlled invariance from Proposition~\ref{prop:charact}. Let $K$ be the maximal robust controlled invariant for the system $\Sigma$ and constraint set $(X,U,D)$ and consider the set $H =\downarrow K$. First, we have from (i) that the set $H$ is a robust controlled invariant for the system $\Sigma$ and constraint set $(X,U,D)$. Moreover, since $K$ is the maximal robust controlled invariant for the system $\Sigma$ and constraint set $(X,U,D)$, one has $H=\downarrow K \subseteq K$. Finally, using the fact that $K \subseteq \downarrow K=H$, one gets $K =\downarrow K$ which implies that $K$ is a lower closed set.

\textbf{\underline{proof of (iii):}} Let $K$ be the maximal robust controlled invariant for the system $\Sigma$ and constraint set $(X,U,D)$ and let $\overline{K}$ be the maximal robust controlled invariant for the system $\Sigma$ and the and constraint set $(X,U,D_{\max})$. First, since $D_{\max} \subseteq D$, we have from Lemma~\ref{lem:inv_cont_dist} that $K \subseteq \overline{K}$. In order to show that $\overline{K} \subseteq K$, and from the maximality of the set $K$ it is sufficient to show that the set $\overline{K}$ is a controlled invariant of the system $\Sigma$ and constraint set $(X,U,D)$, and which is equivalent, from Proposition~\ref{prop:charact} to the following condition:
\begin{equation}
    \forall x \in \overline{K},~ \exists u\in U~ \text{ s.t }~ f(x,u,D) \subseteq \overline{K}.
\end{equation}
Consider $x \in \overline{K}$, we have the existence of $u \in U$ such that $f(x,u,D_{\max}) \subseteq \overline{K}$. Moreover, since the set of disturbance inputs $D$ is closed and bounded above and using the fact that $\Sigma$ is DSM, one has from (iii) in Proposition~\ref{prop:characterizations_monotone} that $f(x,u,D)= f(x,u,\downarrow D_{\max}) \subseteq \downarrow f(x,u,D_{\max})$. Hence, one gets that $f(x,u,D) \subseteq \downarrow f(x,u,D_{\max}) \subseteq \downarrow \overline{K}= \overline{K}$, where the last equality follows from (i). Hence $\overline{K} \subseteq K$ and (iii) holds.

% Now consider $x \in \overline{K}$ and let us show that $x \in K$. From definition of the set $\overline{K}$ there exists a controller $\kappa:X \rightarrow U$ such that for any disturbance input $\mathbf{d}_{\max}:\mathbb{N}_{\geq 0} \rightarrow D_{\max}$, we have $\Phi_\kappa(k,x,\mathbf{d}_{\max}) \in K$ for all $k\in \mathbb{N}_{\geq 0}$. Consider a disturbance input $\mathbf{d}:\mathbb{N}_{\geq 0} \rightarrow D$, using the fact that the set of disturbance inputs $D$ is closed and bounded above we have the existence of $\mathbf{d}':\mathbb{N}_{\geq 0} \rightarrow D_{\max}$ such that $\mathbf{d} \leq_{\mathcal{D}^w} \mathbf{d}'$, which implies from Corollary~\ref{coro:cont_mono} that $\Phi_\kappa(k,x,\mathbf{d}) \leq_{\mathcal{X}} \Phi_\kappa(k,x,\mathbf{d}') \in \overline{K} \subseteq X$, for all $k \in \mathbb{N}_{\geq 0}$. Hence, from the lower closedeness of $X$, one has $\Phi_\kappa(k,x,\mathbf{d}) \in X$ for all $k\in \mathbb{N}_{\geq 0}$. It follows then from maximality of the robust controlled invariant $K$ that $x \in K$. Hence, $K=\overline{K}$ and (iii) holds.  

\textbf{\underline{proof of (iv):}} 
Let $K$ be the maximal robust controlled invariant for the system $\Sigma$ and constraint set $(X,U,D)$ and let $\underline{K}$ be the maximal robust controlled invariant for the system $\Sigma$ and the constraint set $(X,U_{\min},D)$. First, since $U_{\min} \subseteq U$, we have from Lemma~\ref{lem:inv_cont_dist} that $\underline{K} \subseteq K$. In order to show that $K \subseteq \underline{K}$, from the maximality of the set $\underline{K}$, it is sufficient to show that the set $K$ is a controlled invariant of the system $\Sigma$ and constraint set $(X,U_{\min},D)$, and which is equivalent, from Proposition~\ref{prop:charact} to the following condition:
\begin{equation}
    \forall x \in K,~ \exists u\in U_{\min}~ \text{ s.t }~ f(x,u,D) \subseteq K.
\end{equation}
Consider $x \in K$, we have the existence of $u \in U$ such that $f(x,u,D) \subseteq K$. Moreover, since the set of control inputs $U$ is closed and bounded bellow we have the existence of $u'\in U_{\min}$ such that $u' \leq_{\mathcal{U}} u$. Since the system $\Sigma$ is CSM, one has from (iii) in Proposition~\ref{prop:characterizations_monotone} that $f(x,u',D) \subseteq \downarrow f(x,\downarrow u,D) \downarrow f(x,u,D) \subseteq \downarrow K=K$, where the last equality follows from (i). Hence $K \subseteq \underline{K}$ and (v) holds.

\textbf{\underline{proof of (v):}} 
The proof is a direct conclusion from (iii), (iv) and the fact that any CDSM system is a CSM and DSM system.

\bigskip

\textbf{\underline{Proof of Proposition~\ref{prop:charac2}:}}

We provide a proof for each item separately.
\textbf{\underline{proof of (i):}} First, it can be easily seen that if the set $K$ is a robust controlled invariant for the system $\Sigma$ and constraint set $(X,U,D)$ then from Proposition~\ref{prop:charact} and using the fact that $\max(K) \subseteq K$ one gets (\ref{eqn:prop_charact1}). Now assume that (\ref{eqn:prop_charact1}) holds and let us show that (\ref{eqn:prop_charact}) holds. Consider $x \in K$, we have the existence of $x' \in \max(K)$ such that $x \leq_{\mathcal{X}} x'$. Then, from (\ref{eqn:prop_charact1}) we have the existence of $u \in U$ such that $f(x',u,D) \subseteq K$. Hence, one gets $f(x,u,D) \subseteq f(\downarrow x',u,D) \subseteq \downarrow f(x',u,D) \subseteq \downarrow K \subseteq K$, where the second inclusion comes from (i) in Proposition~\ref{prop:characterizations_monotone} and the last inclusion comes from (ii) in Theorem~\ref{thm1}. Hence, condition (\ref{eqn:prop_charact}) holds, and the set $K$ is a robust controlled invariant for the system $\Sigma$ and the constraint set $(X,U,D)$.

\textbf{\underline{proof of (ii):}} From (i) and since the system $\Sigma$ is DSM, to show (ii), it is sufficient to show the equivalence between conditions (\ref{eqn:prop_charact1}) and (\ref{eqn:prop_charact2}). Since $D_{\max} \subseteq D$, one gets directly that (\ref{eqn:prop_charact1}) implies (\ref{eqn:prop_charact2}). Let us show the converse result, consider $x \in \max(K)$, from (\ref{eqn:prop_charact2}) one has the existence of $u \in U$ such that $f(x,u,D_{\max}) \subseteq K$. Hence, one gets that $f(x,u,D) \subseteq f(x,u,\downarrow D_{\max}) \subseteq \downarrow f(x,u,D_{\max}) \subseteq \downarrow K \subseteq K$, where the first inclusion comes from the fact that $D \subseteq \downarrow D_{\max}$, the second inclusion comes from (iii) in Proposition~\ref{prop:characterizations_monotone} and the last inclusion comes from (iii) in Theorem~\ref{thm1}. Hence, condition (\ref{eqn:prop_charact1}) holds.

\textbf{\underline{proof of (iii):}} From (i) and since the system $\Sigma$ is CSM, to show (iii), it is sufficient to show the equivalence between conditions (\ref{eqn:prop_charact1}) and (\ref{eqn:prop_charact3}). Since $U_{\min} \subseteq U$, one gets directly that (\ref{eqn:prop_charact3}) implies (\ref{eqn:prop_charact1}). Let us show the converse result, consider $x \in \max(K)$, from (\ref{eqn:prop_charact1}) one has the existence of $u \in U$ such that $f(x,u,D) \subseteq K$. Moreover, we have the existence of $u' \in U_{\min}$ such that $u' \leq_{\mathcal{U}} u$. Hence, one gets that $f(x,u',D) \subseteq f(x,\downarrow u, D) \subseteq \downarrow f(x,u,D) \subseteq \downarrow K \subseteq K$, where the second inclusion comes from (ii) in Proposition~\ref{prop:characterizations_monotone} and the last inclusion comes from (iv) in Theorem~\ref{thm1}. Hence, condition (\ref{eqn:prop_charact1}) holds.

\textbf{\underline{proof of (iv):}} 
The proof is a direct conclusion from (ii), (iii) and the fact that any CDSM system is a CSM and DSM system.

\bigskip

\textbf{\underline{Proof of Proposition~\ref{prop:feas_inv}:}}

We only provide a proof of (i), the proof of (ii) can be derived similarly. Assume that $x_0$ is closed loop lower feasible w.r.t the constraint set $(X,U,D)$, hence then there exist a controller $\kappa:X \rightarrow U$ and $N \in \mathbb{N}_{>0}$ such that conditions (\ref{eqn:feas11}) and (\ref{eqn:feas12}) are satisfied.
To show the result we use the characterization of controlled invariants in Proposition~\ref{prop:charact}. Consider $x \in K$, we have the existence of $p \in \{0,1,\ldots,N-1\}$ such that $x \in \downarrow \Phi_{\kappa}(p,x_0,D)$. Hence, there exists $\mathbf{d}:\mathbb{N}_{\geq 0} \rightarrow D$ such that $x \leq_{\mathcal{X}} \Phi_{\kappa}(p,x_0,\mathbf{d})$. Consider $u=\kappa(\Phi_{\kappa}(p,x_0,\mathbf{d}))$ and any $d=\mathbf{d}(p)\in D$, using the fact that the system $\Sigma$ is SM we have $f(x,u,d) \in \downarrow f(\Phi_{\kappa}(p,x_0,\mathbf{d}),\kappa(\Phi_{\kappa}(p,x_0,\mathbf{d})),\mathbf{d}(p))= \downarrow \Phi_{\kappa}(p+1,x_0,\mathbf{d})$. Hence, we have two cases
\begin{itemize}
    \item If $p \in \{0,1,\ldots,N-2\}$, then one has  $f(x,u,d) \subseteq  \downarrow \Phi_{\kappa}(p+1,x_0,\mathbf{d}) \subseteq  \downarrow \bigcup\limits_{0 \leq k \leq N-1} \Phi_{\kappa}(k,x_0,D) =K$ .
    \item If $p=N-1$, then one has from (\ref{eqn:feas12}) that $f(x,u,d) \subseteq  \downarrow \Phi_{\kappa}(p+1,x_0,\mathbf{d}) \subseteq \downarrow \Phi_{\kappa}(N,x_0,D) \subseteq  \downarrow \bigcup\limits_{0 \leq k \leq N-1} \Phi_{\kappa}(k,x_0,D) =K$
\end{itemize}
Hence, it follows from Proposition~\ref{prop:charact} that the set $K$ is a controlled invariant for the system $\Sigma$ and constraint set $(X,U,D)$.

\bigskip

\textbf{\underline{Proof of Lemma~\ref{lem:DSM_feas}:}}

Consider $x \in \Phi(k,x_0,\mathbf{u},D)$, we have the existence of $\mathbf{d}:\mathbb{N}_{\geq 0} \rightarrow D$ such that $x=\Phi(k,x_0,\mathbf{u},\mathbf{d})$. Moreover, we have the existence of $\mathbf{d}_{\max}:\mathbb{N}_{\geq 0} \rightarrow D_{\max}$ such that $\mathbf{d} \leq_{\mathcal{D}^w} \mathbf{d}_{\max}$. Then, using the fact that the system $\Sigma$ is DSM, one has $x= \Phi(k,x_0,\mathbf{u},\mathbf{d}) \leq \Phi(k,x_0,\mathbf{u},\mathbf{d}_{\max}) \in \Phi(k,x_0,\mathbf{u},D_{\max})$, for all $k \in \mathbb{N}_{\geq 0}$, which implies from Lemma~\ref{lem:mono} that $\Phi(k,x_0,\mathbf{u},D) \subseteq \downarrow \Phi(k,x_0,\mathbf{u},D_{\max})$, for all $k \in \mathbb{N}_{\geq 0}$.

\bigskip

\textbf{\underline{Proof of Proposition~\ref{prop:feas}:}}

\textbf{\underline{Necessary condition:}} From the open-loop feasibility of $x_0$ w.r.t the constraint set $(X,U,D)$ we have the existence of a control input $\mathbf{u}:\mathbb{N}_{\geq 0} \rightarrow U$ and $N \in \mathbb{N}_{>0}$ such that (\ref{eqn:feas1o}) and (\ref{eqn:feas2o}) hold. Using (\ref{eqn:feas1o}) and the fact that $D_{\max} \subseteq D$, we have that $\Phi(k,x_0,\mathbf{u},D_{\max}) \subseteq \Phi(k,x_0,\mathbf{u},D) \subseteq X$ for all $0 \leq k \leq N-1$. Let us show that the second condition holds, we have
 \begin{align*}
   \Phi(N,x_0,\mathbf{u},D_{\max}) &\subseteq \Phi(N,x_0,\mathbf{u},D)\\ &\subseteq \downarrow \bigcup\limits_{0 \leq k \leq N-1} \Phi(k,x_0,\mathbf{u},D)\\ &\subseteq \downarrow \bigcup\limits_{0 \leq k \leq N-1} \Phi(k,x_0,\mathbf{u},D_{\max})
 \end{align*}
where the first inclusion follows from the fact that $D_{\max} \subseteq D$, the second inclusion comes from (\ref{eqn:feas2o}) and the last inclusion comes from Lemma~\ref{lem:DSM_feas}. 

\textbf{\underline{Sufficient condition:}} From the feasibility of $x_0$ w.r.t the constraint set $(X,U,D_{\max})$ we have the existence of a control input $\mathbf{u}:\mathbb{N}_{\geq 0} \rightarrow U$ and $N \in \mathbb{N}_{>0}$ such that the following conditions are satisfied
\begin{equation}
    \label{eqn:feaspr1}
    \Phi(k,x_0,\mathbf{u},D_{\max}) \subseteq X, \quad \forall~ k \in \{1,2,\ldots,N-1\}
\end{equation}
and 
\begin{equation}
    \label{eqn:feaspr2}
    \Phi(N,x_0,\mathbf{u},D_{\max}) \subseteq \downarrow \bigcup\limits_{0 \leq k \leq N-1} \Phi(k,x_0,\mathbf{u},D_{\max})
\end{equation}
First, we have $\Phi(k,x_0,\mathbf{u},D) \subseteq  \Phi(k,x_0,\mathbf{u},\downarrow D_{\max}) \downarrow  \Phi(k,x_0,\mathbf{u},D_{\max}) \subseteq \downarrow X=X$, for all $0 \leq k \leq N-1$, where the first inequality comes from Lemma~\ref{lem:DSM_feas}, the second inequality follows from (\ref{eqn:feaspr1}) and the last inequality comes from the lower closedeness of the set $X$. To show that (\ref{eqn:feas2o}) holds, we have the following 
\begin{align*}
    \Phi(N,x_0,\mathbf{u},D) &\subseteq \downarrow \Phi(N,x_0,\mathbf{u},D_{\max})\\ &\subseteq  \downarrow \bigcup\limits_{0 \leq k \leq N-1} \Phi(k,x_0,\mathbf{u},D_{\max})\\ &\subseteq \downarrow \bigcup\limits_{0 \leq k \leq N-1} \Phi(k,x_0,\mathbf{u},D)
    \end{align*}
where the first inclusion comes from Lemma~\ref{lem:DSM_feas}, the second inclusion comes from the fact that $x_0$ is feasible w.r.t the constraint set $(X,U,D_{\max})$ and the last inclusion follows from the fact that $D_{\max} \subseteq D$.

\bigskip

\textbf{\underline{Proof of Proposition~\ref{prof:feas_CSM}:}}

Consider $x_0 \in X$, first since $U_{\min} \subseteq U$ one directly has that $x_0$ is open-loop feasible w.r.t the constraint set $(X,U,D)$ if $x_0$ is open-loop feasible w.r.t the constraint set $(X,U_{\min},D)$. Let us show the second implication. Since $x_0$ is open-loop feasible w.r.t the constraint set $(X,U,D)$, one gets the existence of an input trajectory $\mathbf{u}:\mathbb{N}_{\geq 0} \rightarrow \mathcal{U}$ and $N \in \mathbb{N}_{>0}$ such that 
\begin{equation}
\label{eqn:pr1}
    \Phi(k,x_0,\mathbf{u},D) \subseteq X, \quad \forall~ k \in \{1,2,\ldots,N-1\}
\end{equation}
and
\begin{equation}
\label{eqn:pr2}
    \Phi(N,x_0,\mathbf{u},D) \subseteq \downarrow \bigcup\limits_{0 \leq k \leq N-1} \Phi(k,x_0,\mathbf{u},D)
\end{equation}
Now consider any input trajectory $\underline{\mathbf{u}}:\mathbb{N}_{\geq 0} \rightarrow \mathcal{U}_{\min}$ such that $\underline{\mathbf{u}} \leq_{\mathcal{U}^w} \mathbf{u}$. Using the fact that the system $\Sigma$ is CDSM and since the set $X$ is lower closed, it follows from (\ref{eqn:pr1}) that for all $k \in \{1,2,\ldots,N-1\}$
\begin{equation}
    \Phi(k,x_0,\underline{\mathbf{u}},D) \subseteq \Phi(k,x_0, \downarrow \mathbf{u},D) \subseteq \downarrow \Phi(k,x_0,  \mathbf{u},D) \subseteq \downarrow X \subseteq   X
\end{equation}
and condition (\ref{eqn:feas11}) in Definition \ref{def:feas} is satisfied.

Now consider $x \in \Phi(N,x_0,\underline{\mathbf{u}},D)$ and let us show that there exists $k\in \{0,\ldots,N-1\}$ and $\Bar{x} \in \Phi(k,x_0,\underline{\mathbf{u}},D)$ such that $x \leq \Bar{x}$. 

For $x \in \Phi(N,x_0,\underline{\mathbf{u}},D)$, we have the existence of a trajectory $\mathbf{d}:\mathbb{N}_{\geq 0} \rightarrow \mathcal{D}$ such that $x=\Phi(N,x_0,\underline{\mathbf{u}},\mathbf{d})$. Moreover, from (\ref{eqn:pr2}) we have for $\Phi(N,x_0,\underline{\mathbf{u}},\mathbf{d})$ the existence of a trajectory $\Bar{\mathbf{d}}:\mathbb{N}_{\geq 0} \rightarrow \mathcal{D}$ with $\mathbf{d} \leq_{\mathcal{D}^w} \Bar{\mathbf{d}}$ and $k \in \{1,\ldots,N-1\}$ such that 
\begin{equation}
\label{eqn:pr22}
\Phi(N,x_0,\mathbf{u},\mathbf{d}) \leq \Phi(k,x_0,\mathbf{u},\Bar{\mathbf{d})}.
\end{equation}
Moreover, since the system $\Sigma$ is CDSM, one has that
\begin{equation}
    \label{eqn:pr3}
    \Phi(k,x_0,\underline{\mathbf{u}},{\mathbf{d})} \leq \Phi(k,x_0,\mathbf{u},{\mathbf{d}}).
\end{equation}
Define $\varepsilon \in \mathbb{R}^n_{\geq 0}$ as $\varepsilon:=\Phi(k,x_0,\mathbf{u},\Bar{{\mathbf{d}}})-\Phi(k,x_0,\underline{\mathbf{u}},\Bar{{\mathbf{d})}}$. We have that 
\begin{align*}
x=&\Phi(N,x_0,\underline{\mathbf{u}},\mathbf{d})  \leq  \Phi(N-k,\Phi(k,x_0,\mathbf{u},{\mathbf{d}}),\underline{\mathbf{u}},{\mathbf{d}})-\varepsilon  \\ &\leq  \Phi(N,x_0,\mathbf{u}, {\mathbf{d}}) - \varepsilon \leq   \Phi(k,x_0,\mathbf{u},\Bar{\mathbf{d}}) - \varepsilon = \Phi(k,x_0,\underline{\mathbf{u}},\Bar{\mathbf{d}})
\end{align*}
where the first follow from the application of (\ref{eqn:feas_CSM_C}) to the inequality in (\ref{eqn:pr3}) $N-k$ times, the second inequality comes from the fact that the system $\Sigma$ is CDSM, the third inequality comes from (\ref{eqn:pr22}) and the last equality comes from the definition of $\varepsilon$. Hence, we have the existence of $k\in \{0,\ldots,N-1\}$ and $\Bar{x}= \Phi(k,x_0,\underline{\mathbf{u}},\Bar{\mathbf{d})} \in \Phi(k,x_0,\underline{\mathbf{u}},D)$ such that $x \leq \Bar{x}$. Hence, condition (\ref{eqn:feas12}) is satisfied and $x_0$ is open-loop feasible w.r.t the constraint set $(X,U_{\min},D)$.

\bigskip

\textbf{\underline{Proof of Proposition~\ref{pro:feas_2}:}}

Assume that $x_0 \in X$ is open-loop feasible w.r.t the constraint set $(X,U,D)$. We have the existence of a control input $\mathbf{u}:\mathbb{N}_{\geq 0} \rightarrow \mathcal{U}$ and $N \in \mathbb{N}_{>0}$ such that (\ref{eqn:feas1o}) and (\ref{eqn:feas2o}) are satisfied. Consider the controller $\kappa:\mathcal{X} \rightarrow \mathcal{U}$ defined as follows: for $k \in \{1,2,\ldots,N-1\}$, $\kappa(\Phi(k,x_0,\mathbf{u},D))=\mathbf{u}(k)$, and $\kappa(\Phi(k,x_0,\mathbf{u},D)) \in \mathcal{U}$ for all $k \geq N$. Using the controller $\kappa$, one can easily check that (\ref{eqn:feas11}) and (\ref{eqn:feas12}) hold. Hence, $x_0$ is closed-loop feasible w.r.t the constraint set $(X,U,D)$. 

Let us now show the converse result under the assumption that $D=\{\overline{d}\}$. Assume that $x_0 \in X$ is closed-loop feasible w.r.t the constraint set $(X,U,D)$. We have the existence of a controller $\kappa:\mathcal{X} \rightarrow \mathcal{U}$ and $N \in \mathbb{N}_{>0}$ such that (\ref{eqn:feas11}) and (\ref{eqn:feas12}). Consider the control input trajectory  $\mathbf{u}:\mathbb{N}_{\geq 0} \rightarrow \mathcal{U}$ defined as follows: for $k \in \{1,2,\ldots,N-1\}$, $\mathbf{u}(k)=\kappa(\Phi(k,x_0,\mathbf{u},\overline{d}))$ and $\mathbf{u}(k) \in \mathcal{U}$ for all $k \geq N$. Using the control input trajectory $\mathbf{u}$, one can easily check that (\ref{eqn:feas1o}) and (\ref{eqn:feas2o}) hold. Hence, $x_0$ is open-loop feasible w.r.t the constraint set $(X,U,D)$. The converse result can also be obtained when the system $\Sigma$ is DSM and $\card(D_{\max})=1$, by using the equivalence between open-loop feasibility w.r.t the constraint sets $(X,U,D)$ and $(X,U,D_{\max})$ for the case of DSM systems, as shown in Proposition~\ref{prop:feas}.

\bigskip

\textbf{\underline{Proof of Theorem~\ref{thm:stric_feas}:}}

Since the set $D$ is compact and $f$ is upper semicontinuous on its first and third arguments, one has that the set valued map $F:X\times U \rightrightarrows X$ defined by $F(x,u):=f(x,u,D)$ is upper semicontinuous on $x$ and for any $u \in U$. Moreover, from continuity of $f$ and compacteness of $D$ we have that $\Phi(N-1,x_0,\mathbf{u},D)$ is compact. Hence, for $\beta_N=\min(\varepsilon_N,\gamma)>0$, where $\gamma>0$ is defined in (ii), we have from Lemma~\ref{lem:app1} the existence of $\varepsilon_{N-1}>0$ such that 
\begin{align*}
    F(\mathcal{B}_{\varepsilon_{N-1}}(\Phi&(N-1,x_0,\mathbf{u},D), \mathbf{u}(N-1))  \\& \subseteq \mathcal{B}_{\beta_N}(F(\Phi(N-1,x_0,\mathbf{u},D), \mathbf{u}(N-1)))\\ &\subseteq \mathcal{B}_{\beta_N}(\Phi(N,x_0,\mathbf{u},D)).
\end{align*}
Let us define the sequences $\varepsilon_k>0$, $k\in \{1,\ldots,N-2\}$ and $\beta_k>0$, $k \in \{1,\ldots,N-1\}$, iteratively as follows: for $k \in \{N-1,N-2,\ldots,1\}$, $\beta_{k}=\min\{\varepsilon_{k},\gamma\}$, where $\gamma>0$ is defined in (ii), and $\varepsilon_{k-1}$ is such that $F(\mathcal{B}_{\varepsilon_{k-1}}(\Phi(k-1,x_0,\mathbf{u}))) \subseteq \mathcal{B}_{\beta_{k}}(F((\Phi(k,x_0,\mathbf{u}))$. The existence of such $\varepsilon_{k-1}$, $k\in \{N-1,N-2,\ldots,1\}$ is guaranteed from Lemma~\ref{lem:app1} using the upper semicontinuity of the map $F$ and the fact that $\Phi(k,x_0,\mathbf{u},D)$ is compact for all $k \in \{1,2,\ldots,N\}$. Hence, one gets
\begin{align}
\label{eqn:thm2}
    \Phi(N,\mathcal{B}_{\beta_0}(x_0),\mathbf{u}) & \subseteq \Phi(N,\mathcal{B}_{\varepsilon_0}(x_0),\mathbf{u}) \nonumber\\ & \subseteq \Phi(N-1,\mathcal{B}_{\beta_1}(\Phi(1,x_0,\mathbf{u}),\mathbf{u}) \nonumber\\ &  \subseteq \Phi(N-1,\mathcal{B}_{\varepsilon_1}(\Phi(1,x_0,\mathbf{u}),\mathbf{u}) \nonumber \\  &\subseteq \Phi(N-2,\mathcal{B}_{\beta_2}(\Phi(2,x_0,\mathbf{u}),\mathbf{u}) \nonumber\\ & \subseteq \Phi(N-2,\mathcal{B}_{\varepsilon_2}(\Phi(2,x_0,\mathbf{u}),\mathbf{u}) \nonumber \\
    &\subseteq \ldots \nonumber
    \\ &\subseteq \Phi(1,\mathcal{B}_{\beta_{N-1}}(\Phi(N-1,x_0,\mathbf{u}),\mathbf{u}) \nonumber\\ & \subseteq \Phi(1,\mathcal{B}_{\varepsilon_{N-1}}(\Phi(N-1,x_0,\mathbf{u}),\mathbf{u}) \nonumber
    \\ &\subseteq \mathcal{B}_{\varepsilon_N}(\Phi(N,x_0,\mathbf{u})) \nonumber \\ &\subseteq \downarrow \bigcup\limits_{0 \leq k \leq N-1} \Phi_{\kappa}(k,x_0,\mathbf{u},D)
\end{align}
where the last inclusion comes from (i). Now let $\beta=\min \beta_i,~i=\{1,2,\ldots,N-1\}$ consider $x_1 \in \{\uparrow x_0\} \cap \mathcal{B}_{\beta}(x_0)$. We have
\begin{align*}
    \Phi(N,x_1,\mathbf{u}) &\subseteq  \Phi(N,\mathcal{B}_{\beta}(x_0),\mathbf{u})\\ &\subseteq  \Phi(N,\mathcal{B}_{\beta_0}(x_0),\mathbf{u})\\ &\subseteq \downarrow \bigcup\limits_{0 \leq k \leq N-1} \Phi_{\kappa}(k,x_0,\mathbf{u},D)\\ &\subseteq \downarrow \bigcup\limits_{0 \leq k \leq N-1} \Phi_{\kappa}(k,x_1,\mathbf{u},D).
\end{align*}
Where the second inclusion comes from (\ref{eqn:thm2}) and the last inclusion follows from the fact that $\Sigma$ is a SM system. Hence, condition (\ref{eqn:feas2o}) of Definition~\ref{def:feas} is satisfied.

Moreover, we have from (ii) that for all $k \in \{0,1,\ldots,N-1\}$
\begin{align*}
    \Phi(k,x_1,\mathbf{u}) &\subseteq \Phi(k,\mathcal{B}_{\beta_{N-k}}(x_0),\mathbf{u})\\ &\subseteq  \mathcal{B}_{\beta_{N-k+1}}(\Phi(k,x_0,\mathbf{u},D))\\ &\subseteq  \mathcal{B}_{\gamma}(\Phi(k,x_1,\mathbf{u},D)) \subseteq X.
\end{align*}
Hence, condition (\ref{eqn:feas2o}) of Definition~\ref{def:feas} is satisfied, and any $x_1 \in \{\uparrow x_0\} \cap \mathcal{B}_{\beta}(x_0)$ is feasible.  
Now, for the case when the map $f$ is $L$-Lipschitz on its first argument, it follows that the set valued map $F:X\times U \rightrightarrows X$ defined by $F(x,u):=f(x,u,D)$ is $L$-lipschitz on its first argument. Hence, the sequences $\varepsilon_k>0$, $k\in \{1,\ldots,N-1\}$ and $\beta_k>0$, $k \in \{1,\ldots,N\}$, can be constructed according to Lemma~\ref{lem:app1} iteratively as follows: for $k \in \{N-1,N-2,\ldots,1\}$, $\beta_{k}=\min\{\varepsilon_{k},\gamma\}$, where $\gamma>0$ is defined in (ii) and $\varepsilon_{k-1}=\frac{\beta_k}{L}$. hence, the result holds with $\beta=\min\beta_i$, $i \in \{1,\ldots,N\}$.

\bigskip

\textbf{\underline{Proof of Proposition~\ref{prop:charac_boundaries}:}}

First, since the map $f:\mathcal{X}\times \mathcal{U}\times \mathcal{D}$ describing the system $\Sigma$ is lower semicontinuous on its first argument and the set of control inputs $U$ and disturbance inputs $D$ are compact, one has from Proposition~\ref{prop:closedness} that $\dom(\kappa)$ is closed and $\partial(\dom(\kappa)) \subseteq \dom(\kappa)$. 

Since $\kappa$ is a maximal invariance controller, it follows directly that $\Phi_{\kappa}(k,x_0,D)\subseteq \dom(\kappa)$ for all $k \in \mathbb{N}_{\geq 0}$. The latter property implies that either $\Phi_{\kappa}(k,x_0,D) \cap \partial(\dom(\kappa)) \neq \emptyset$ for all $k \in \mathbb{N}_{\geq 0}$, which is consistent with (i), or there exists $p \in \mathbb{N}_{>0}$ such that $\Phi_{\kappa}(p,x_0,D) \subseteq \Int(\dom(K))$. Assume w.l.o.g that $p$ is the minimal integer such that $\Phi_{\kappa}(p,x_0,D) \subseteq \Int(\dom(\kappa))$. Let us show, by contradiction, that (ii) holds. Assume that $\Phi_{\kappa}(p-1,x_0,D) \cap (\partial(\dom(\kappa))\setminus \partial X) \neq \emptyset$ and let $z \in \Phi_{\kappa}(p-1,x_0,D) \cap (\partial(\dom(\kappa))\setminus \partial X)$. Since $\Int(\dom(\kappa))$ is open, we have the existence of $\varepsilon >0$ such that $\mathcal{B}_{\varepsilon}(\Phi_{\kappa}(p,x_0,D)) \subseteq \Int(\dom(\kappa))$. Moreover, since the set $D$ is compact and $f$ is upper semicontinuous on its first and third arguments, one has that the set valued map $F:X\times U \rightrightarrows X$ defined by $F(x,u):=f(x,u,D)$ is upper semicontinuous on $x$. Since $z \in \partial(\dom(\kappa))$, for $\varepsilon>0$ and from the upper semicontinuity of $F$ we have the existence of an $\eta>0$ such that
\begin{align*}
   F(\mathcal{B}_{\eta}(z),\kappa(z)) & \subseteq \mathcal{B}_{\varepsilon}(F(z,\kappa(z)))\\ &\subseteq \mathcal{B}_{\varepsilon}(F(\Phi_{\kappa}(p-1,x_0,D),\kappa(z)))\\ &= \mathcal{B}_{\varepsilon}(f(\Phi_{\kappa}(p-1,x_0,D),\kappa(z),D))\\ &=\mathcal{B}_{\varepsilon}(\Phi_{\kappa}(p,x_0,D)) \subseteq \Int(\dom(\kappa)). 
\end{align*}
Hence, there exists $z' \in (X\setminus \dom(\kappa)) \cap \mathcal{B}_{\eta}(z)$ and $u'=\kappa(z) \in U$ such that 
\begin{align}
    F(z',u')=f(z',\kappa(z),D) \subseteq F(\mathcal{B}_{\eta}(z),\kappa(z)) \subseteq \Int(\dom(\kappa)).
\end{align}
The last property contradicts the maximality of the set $\dom(\kappa)$. Hence, we have necessarily that $\Phi_{\kappa}(p-1,x_0,D) \cap (\partial(\dom(\kappa))\setminus \partial X)= \emptyset$. Finally, since $\Phi_{\kappa}(p-1,x_0,D) \subseteq \dom(\kappa)$ and $\Phi_{\kappa}(p-1,x_0,D) \nsubseteq \Int(\dom(\kappa))$, one gets $\Phi_{\kappa}(p-1,x_0,D) \cap \partial X \neq \emptyset$, which implies (ii).

\bigskip

\textbf{\underline{Auxiliary result:}}
\begin{lemma}
\label{lem:app1}
Consider an upper semicontinuous set-valued map $F: \mathcal{X} \rightrightarrows \mathbb{R}^n$ and consider a compact set $Z \subseteq \mathcal{X}$. For all $\varepsilon >0$, there exists $\eta>0$ such that $F(\mathcal{B}_{\eta}(Z)\cap \mathcal{X}) \subseteq \mathcal{B}_{\varepsilon}(F(Z))$. Moreover, when the map $F$ is $L$-Lipschitz on $\mathcal{X}$, for $L>0$, then the property holds for any $\eta \leq \varepsilon/L$.
\end{lemma}
\textbf{{\it Proof: }}
Consider $\varepsilon > 0$ and $x \in Z$. Since $F$ is upper semicontinuous, we have the existence of $\eta_x>0$ such that $F(\mathcal{B}_{\eta_x}(x)) \cap \mathcal{X}) \subset \mathcal{B}_{\varepsilon}(F(x))$. Let $\eta=\min_{x \in Z}\eta_x$. It follows from the compacteness of the set $Z$ that $\eta>0$. Hence, one gets:
\begin{align*}
F(\mathcal{B}_{\eta}(Z)\cap \mathcal{X}) &= \bigcup\limits_{x \in Z} F(\mathcal{B}_{\eta}(x)\cap \mathcal{X})\\ &\subseteq \bigcup\limits_{x \in Z} F(\mathcal{B}_{\eta_x}(x)\cap \mathcal{X})\\ &\subseteq \bigcup\limits_{x \in Z} \mathcal{B}_{\varepsilon}(F(x))\\ &=\mathcal{B}_{\varepsilon}(F(Z)).
\end{align*}
The last result follows immediately from the fact that the map $F$ is $L$-Lipschitz, since 
$$F(\mathcal{B}_{\frac{\varepsilon}{L}}(Z) \cap \mathcal{X}) \subseteq \mathcal{B}_{\varepsilon}(F(Z)).$$

\end{appendices}

\vspace{1cm}

\bmhead{Acknowledgments}

This work was supported by the ANR PIA funding: ANR-20-IDEES-0002, by the NSF grants CNS-2111688, ECCS-1906164, and AFOSR grant FA9550-21-1-0288.

\bmhead{Conflict of interest} 
The authors declare no conflict of interest related to this work.


\bibliographystyle{abbrv}
%\bibliographystyle{unsrt}
%\bibliography{ref}

\bibliography{manuscript.bib}% common bib file


\end{document}